\documentclass{article}

\pdfoutput=1
\usepackage{url}
\usepackage{hyperref}
\usepackage[english]{babel}

\usepackage{graphicx} 
\usepackage{color}

 \usepackage[table,xcdraw]{xcolor} 
 
\usepackage{amsmath}

\def\R{ {\rm \,I\!R} }    
\def\Sum#1#2{\sum\limits_{#1}^{#2}}
\def\fracg#1#2{{\displaystyle{\frac{#1}{#2}}}}  

\title{Data-Driven Methods to Monitor, Model, Forecast and Control Covid-19 Pandemic: Leveraging Data Science, Epidemiology and Control Theory}

\author{Teodoro Alamo\footnote{Departamento de Ingenier\'ia de Sistemas y Autom\'atica, Universidad de Sevilla, Escuela Superior de Ingenieros, Camino de los Descubrimientos s/n, 41092 Sevilla, Spain (e-mail: talamo@us.es) }, Daniel G. Reina\footnote{Departamento de Ingenier\'ia Electrónica, Universidad de Sevilla, Escuela Superior de Ingenieros, Camino de los Descubrimientos s/n, 41092 Sevilla, Spain (e-mail: dgutierrezreina@us.es) }, Pablo Millán Gata \footnote{Departamento de Ingeniería, Universidad Loyola Andalucía, 41014 Seville, Spain (e-mail: pmillan@uloyola.es)}}

\date{}

\begin{document}

\maketitle
\thispagestyle{empty}

\begin{abstract}
This document analyzes the role of data-driven methodologies in Covid-19 pandemic. We provide a SWOT analysis and a roadmap that goes from the access to data sources to the final decision-making step. We aim to review the available methodologies while anticipating the difficulties and challenges in the development of data-driven strategies to combat the Covid-19 pandemic. A 3M-analysis is presented: Monitoring, Modelling and Making decisions. The focus is on the potential of well-known data-driven schemes to address different challenges raised by the pandemic: i) monitoring and forecasting the spread of the epidemic; (ii) assessing the effectiveness of government decisions; (iii) making timely decisions. Each step of the roadmap is detailed through a review of consolidated theoretical results and their potential application in the Covid-19 context. When possible, we provide examples of their applications on past or present epidemics. We do not provide an exhaustive enumeration of methodologies, algorithms and applications. We do try to serve as a bridge between different disciplines required to provide a holistic approach to the epidemic: data science, epidemiology, control-theory, etc. That is, we highlight effective data-driven methodologies that have been shown to be successful in other contexts and that have potential application in the different steps of the proposed roadmap. To make this document more functional and adapted to the specifics of each discipline, we encourage researchers and practitioners to provide feedback\footnote{conco.team@gmail.com}. We will update this document regularly.     

{\bf CONCO-Team:} The authors of this paper belong to the CONtrol COvid-19 Team, which is composed of more than 35 researches from universities of Spain, Italy, France, Germany, United Kingdom and Argentina. The main goal of CONCO-Team is to develop data-driven methods for the better understanding and control of the pandemic. 

\end{abstract}


\newpage

\tableofcontents
\newpage

\bibliographystyle{plain}

\section{Introduction}

The outbreak of 2019 novel coronavirus disease (Covid-19) is a public health emergency of international concern. Governments, public institutions, health-care professionals and researches of different disciplines are addressing the problem of controlling the spread of the virus while reducing the negative effect on the economy and society. The challenges raised by the pandemic require a holistic approach. In this document we analyze the interplay between data science, epidemiology and control theory in the decision making process. In line with the current and urgent needs identified by epidemiologists \cite{Gog2020}, this paper aims to sift the available methodologies while anticipating the difficulties and challenges encountered in the development of data-driven strategies to combat the Covid-19 pandemic. Data-driven schemes can be fundamental to: i) monitor, model and forecast the spread of the epidemic; (ii) assess the potential impacts of government decisions not only from a health-care point of view but also from an economic and social one; (iii) make timely decisions.

Data-driven community is formed by those researches and practitioners that develop prediction models and decision-making tools based on data. As an initial step previous to the description of the available methodologies, the strengths, weaknesses, opportunities and threats encountered by this community when addressing the multiple challenges raised by Covid-19 pandemic are discussed by means of a  SWOT analysis. 

\begin{figure}[ht!]
    \centering
    \includegraphics[width=1.0\textwidth]{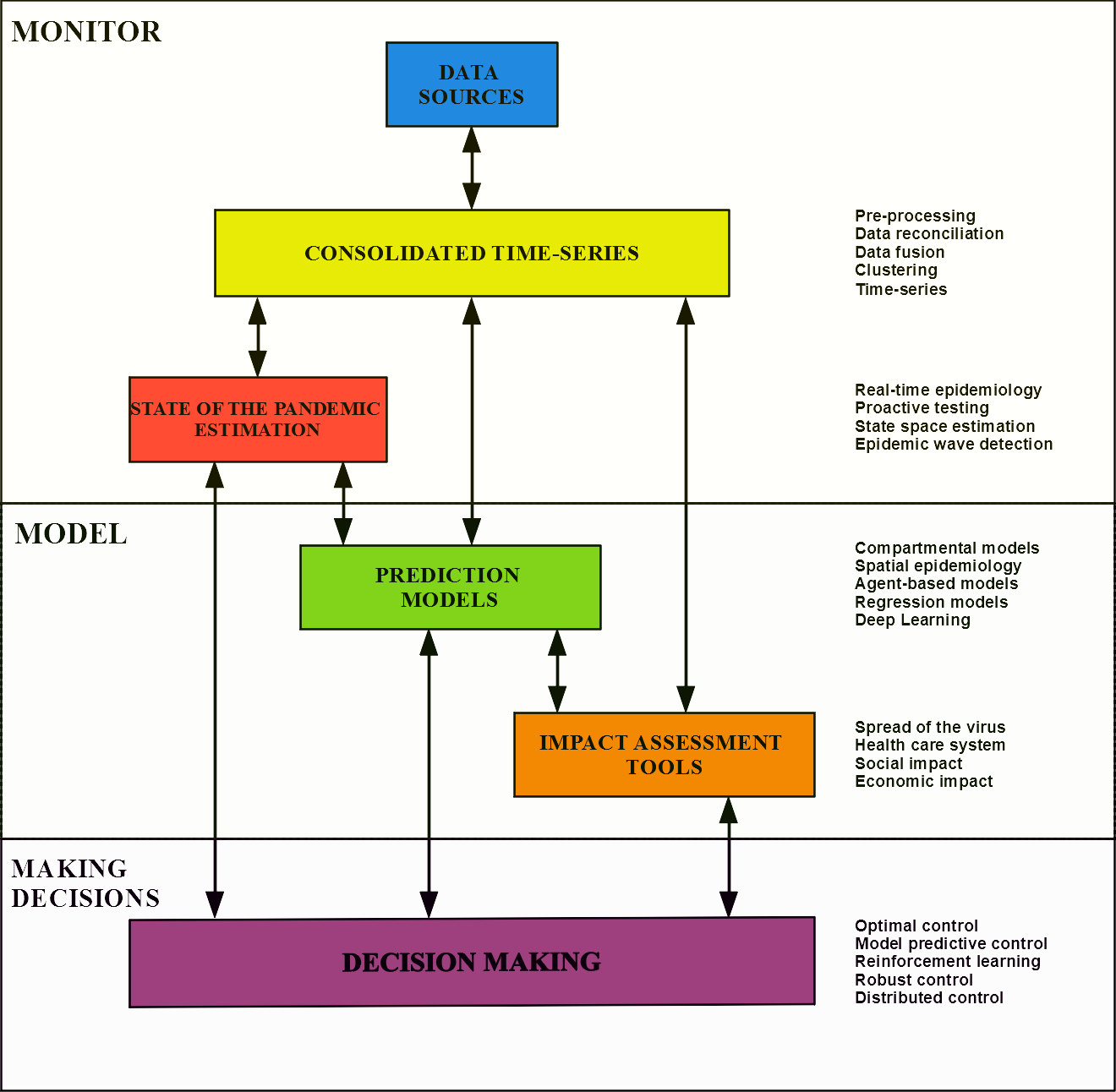}
    \caption{3M-Approach to data-driven control of an epidemic: Monitoring, Modelling and Making Decisions.}
    \label{fig:structure}
\end{figure}

Optimal decision making in the context of Covid-19 pandemic is a complex process that requires to deal with a significant amount of uncertainty and the severe consequences of not reacting timely and with the adequate intensity. In this document, a roadmap that goes from the access to data sources to the final decision-making step is provided. A 3M-analysis is proposed: Monitoring, Modelling and Making decisions. See Figure \ref{fig:structure}.
Each step of the roadmap is analyzed through a review of consolidated theoretical results and their potential use in the Covid-19 context.   When possible, examples of applications of these methodologies on past or present epidemics are provided. Data-driven methodologies that have been shown to be successful in other biological contexts (e.g \cite{Maddalena2020}), or that have been identified as promising solutions in the present pandemic, are highlighted. This document does not provide an exhaustive enumeration of methodologies, algorithms and applications. Instead, it is conceived to serve as a bridge between the different disciplines required to provide a holistic approach to the epidemic: data science, epidemiology and control theory. 

Data is a fundamental pillar to understand, model, forecast, and manage many of the aspects required to provide a comprehensive response. There exists many different open data resources and institutions providing relevant information not only in terms of specific epidemiological Covid-19 variables, but also of other auxiliary variables that facilitate the assessment of the effectiveness of the implemented interventions. See \cite{Alamo2020b} for a review on Covid-19 open data resources and repositories.  

\vspace{2mm}

Data reconciliation techniques play a relevant role in the proposed approach since the available data sources suffer from severe limitations. Methodologies like data reconciliation, data-fusion, data-clustering, signal processing, to name just a few, can be used to detect anomalies in the raw data and generate time-series with enhanced quality.  

Another important aspect on the 3M-approach is the real-time surveillance of the epidemic. This is implemented by monitorization of the mobility, the use of social media to assess the compliance of the restrictions and recommendations, pro-active testing, contact-tracing, etc. In this context is also relevant the design and implementation of surveillance systems capable of detecting secondary waves of the pandemic.  

Modelling techniques are called to play a relevant role in the combat against Covid-19 \cite{Rhodes2020:MODEL:MATHS}. Epidemiological models range from low dimensional compartmental models to complex spatially distributed ones. Fundamental parameters that characterize the spreading capacity of the virus can be obtained from the adjusted models. Besides, data-driven parametric inference provides mechanisms to anticipate the effectiveness of the adopted interventions. However, fitting the models to the available data requires specific techniques because of critical issues like partial observation, non-linearities and non-identifiability. Sensitivity analysis, model selection and validation methodologies have to be implemented.  
Apart from the forecasting possibilities that epidemiological models offer, there exists other possibilities. Different forecasting techniques from the field of data science can be applied in this context. The choice ranges from simple linear parametric methods to complex deep-learning approaches. The methods can be parametric or non parametric in nature. Some of these techniques provide probabilistic characterizations of the provided forecasts.  

There are a myriad of potential measures to mitigate the epidemic, but some might not be effective \cite{Xiao2020:CONTROL:MEASURES}. Besides, some of them, like lock-down of an entire country, have an unbearable effect on economy and should be adopted at the precise moment and for the shortest period of time. Others, like pro-active testing and contact-tracing can be very effective and have a minor impact on the economy \cite{Ferretti2020}. Control theory provides a consolidated framework to formulate many of the decision-making problems: optimal allocation of resources, determination of the optimal moment to strengthen the mitigation interventions, etc. The use of optimal control theory in epidemic control has a long history. We also mention the potential of (distributed) model predictive control. Control theory, along with other mathematical methodologies like bifurcation theory, lyapunov theory, etc. have been extensively used to characterize the different possible qualitative behaviours of a given epidemic. 

This document is organized as follows. In Section \ref{sec:Swot} we provide a SWOT analysis of the role of data-driven approaches in the control of the Covid-19 pandemic. Section \ref{sec:data:sources} describes the main data sources that can be used to develop data-driven methods. 
Section \ref{Sec:Consolidated:Time:Series} is devoted to the available methodologies to improve the quality of data and the generation of consolidated time-series. Section \ref{sec:Estimation:State:Epidemic} describes different methodologies to monitor the current state of the pandemic. An overview of the different techniques to model the epidemic is provided in Section \ref{sec:models}. The main forecasting techniques are described in Section \ref{sec:forecasting}. 
The question of how to assess the effectiveness of different non-pharmaceutical measures is analyzed in Section \ref{sec:Impact:Tools}. 
The decision making process, and its link with control theory is addressed in Section \ref{sec:Decision:Making}. The paper is finished with a section of conclusions.

\section{SWOT analysis of data-driven methods in Covid-19 pandemic}\label{sec:Swot}

\begin{figure}[ht!]
    \centering
    \includegraphics[width=0.8\textwidth]{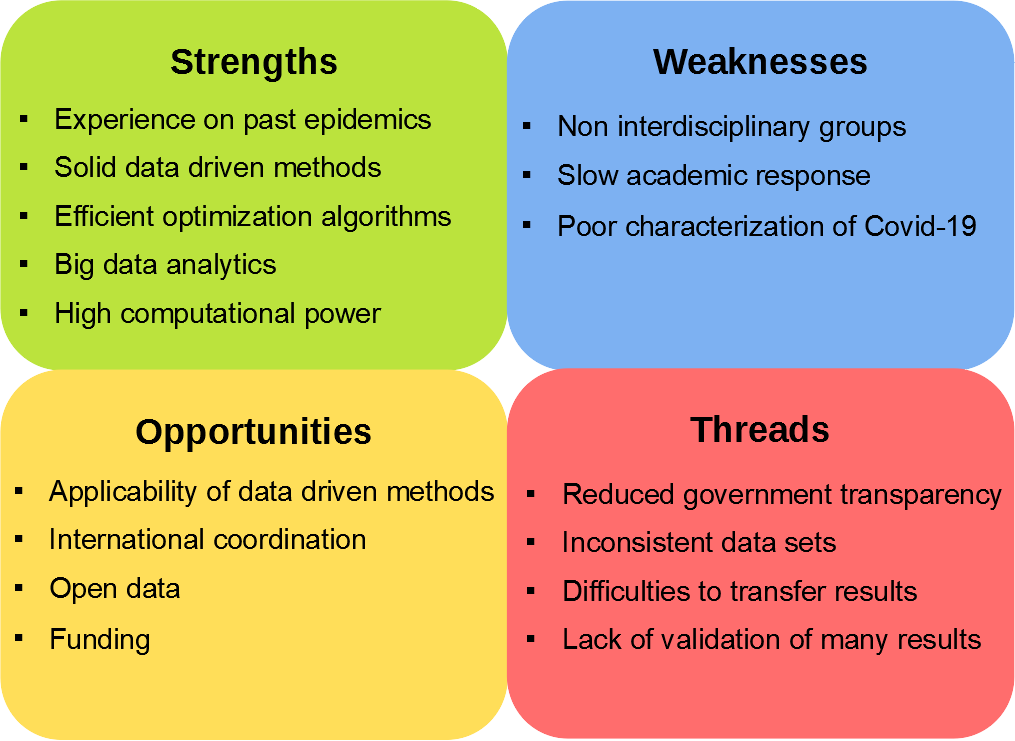}
    \caption{SWOT}
    \label{fig:swot}
\end{figure}

With the aim of providing an overview of the potential impact of data-driven methodologies in the control of Covid-19 pandemic, a SWOT analysis identifying strengths, weaknesses, opportunities, and threats is presented in this section. A summary of the conducted SWOT analysis is presented in Figure \ref{fig:swot}. In the following subsections, the identified bullet points are further developed.

\vspace{2mm}

\subsection{Strengths}
Some of the main strengths of data-driven methods and related research groups fighting Covid-19 are given below: 

\begin{itemize}
\item\textbf{Experience acquired on past epidemics:} Recent epidemics preceding Covid-19, like SARS or MERS, motivated a huge amount of research in the past (see for example \cite{Heesterbeek2015}, \cite{Alvarez2015:MORILLA} and the references therein). This previous research effort not only provides invaluable information about other epidemics, but it also enables today's researches to count on data-driven techniques developed to estimate, model, forecast, and make decisions in the context of an epidemic outbreak. For example, \cite{Brauer2000} contains a comprehensive collection of mathematical models for epidemiology. In \cite{Wu2020} relevant information about previous coronavirus epidemics is presented.

\item\textbf{Solid theoretical foundations of data-driven methods:} Data-driven methods are supported by strong theoretical foundations. This enables decision-makers to manage pandemics with different tools, adapted to different contexts and data, and guaranteeing different degrees of certainty and efficacy.

\item\textbf{Efficient optimization algorithms and solvers:} Many data-driven methods are formulated as the solution of one or several optimization problems, which can be addressed by means of efficient optimization algorithms and solvers \cite{Bottou2018}.  

\item\textbf{Big data analytics resources:} Accurate and effective tools to control the evolution of a pandemic require gathering and processing pervasive data. Big data analytics, developed in the 21st century to an unprecedented level, provides hundreds of different information  systems. Some examples are the use of mobile phones for contact tracing \cite{Rao2020} or real-time mapping of epidemics using social networks \cite{COVIDcast2020}.

\item \textbf{High computation capacity}: The continuous fulfilment of Moore's law for more than 50 years has made possible enormous advances in hardware technologies, such as supercomputers, clusters, and cloud computing, that include a large number of processors and graphical processing units (GPUs). Such computation capacity benefits the development of complex data-driven methods.

\end{itemize}

\subsection{Weaknesses}
In what regards effectiveness fighting Covid-19, some general flaws of the academic community are detailed next:
\begin{itemize}
\item\textbf{Many research groups do not possess the required interdisciplinary:} The works reviewed in this survey suggest a relevant lack of interdisciplinary. Many reported analysis and results are not conducted joining efforts of, for example, epidemiologists, data-science scientists, experts in system engineering and economists. This is likely to produce results and recommendations that might be biased or may tend to overlook aspects related to public health, advanced statistic tools, dynamics effects or economic impact.
\item\textbf{A solid and validated academic response is often slow:} Academic outcomes come often in the form of academic publications and tools to produce predictions and recommendations. Solid academic results require time to collect and process reliable data, make developments, conduct validation and go through a peer-reviewing process.
\item\textbf{Poor characterization of Covid-19:} Although a huge amount of effort has been done to determine the main characteristics of Covid-19, months after the epidemic outbreak, it is still difficult to count on consolidated results. As an example, there are still many open questions related to the seasonal behaviour of the virus or the duration of the immunological protection after recovery. 
\end{itemize}

\subsection{Opportunities}

Some of the most important opportunities are detailed below:

\begin{itemize}

\item\textbf{Applicability of data-driven methodologies:} As it is detailed in this document, many epidemiology subproblems, ranging from the estimation of the epidemic characteristics to forecasting and assessment of government measures, can be addressed using data-driven techniques. Decisions taken at the right time, like partial/total lock-downs, can save thousands of lives while limiting the damage to other socio-economic aspects. A quite comprehensive review of the measures taken by 11 European countries is made in \cite{Flaxman2020}, estimating a total impact in the reduction of the number of deaths from 87.000 to 29.000, only for Europe and up until 30 March 2020.

\item\textbf{Coordinating institutions:} A relevant number of institutions, like the World Health Organization and the different Centers for Disease Prevention and Control from a national or continental scope, are making a great effort to provide a coordinated response to the epidemic. Furthermore, many governments and research institutions have created interdisciplinary task forces aimed at developing data-driven approaches to fight Covid-19. 

\item\textbf{Availability of many open data sources:} The increasing number of institutions and research teams working against Covid-19 is providing an invaluable amount of meaningful, open-data resources to address the pandemic from a data science point of view (see \cite{alamo2020open} for an actualized survey of the main institutions and open-data repositories to fight Covid-19).

\item\textbf{Funding:}
The unthinkable social and economic impact of the Covid-19 pandemic is fostering the mobilization of huge public and private economic resources for related research.
\end{itemize}

\subsection{Threats}

Finally, this section summarizes the most important threats to success fighting the virus with data-driven tools:

\begin{itemize}

\item\textbf{Reduced government transparency:} The secrecy on many aspects of the pandemic of some governments is hindering the access to valuable information \cite{HoldenThorp2020}.

\item\textbf{Inconsistent data sets:} Unfortunately, the quality of the available data is far from ideal because of a good number of issues like changing criteria, insufficiently documented large diversity of
sources and formats, non-comparable metrics between countries,  aggregated data without a clear timestamp (often, some significant increases in the time series are due to the aggregation on several days), etc. \cite{Alamo2020b}.

\item\textbf{Difficulties to transfer obtained results to society:} The complexity of pandemic evolution, aggravated in today's highly entangled and global world, makes it complex to transfer meaningful results to society in a clear way.

\item\textbf{Lack of validation of many results:}
The complexity of the phenomena and the need for a rapid response against Covid-19 involves important risks. Many of the published results suffer from a lack of validation or test to assess their performance. Therefore, results or recommendations based on insufficiently corroborated analysis may be transferred to society at a given moment. This could explain, for example, the discrepant recommendations on the use of masks given by different national and international health institutions \cite{Eikenberry2020}.
\end{itemize}

\section{Data sources}\label{sec:data:sources}
Open data resources play an important role in the fight against Covid-19. Time series of the number of confirmed cases and deaths rates, among other indicators, are daily analyzed by the scientific community. The objective is to study the spread and impact of Covid-19, both worldwide and locally in each country or region, to evaluate the impact on several aspects such as citizens life, health systems, and economy. The main open data resources related to Covid-19 are summarized in \cite{alamo2020open} and \cite{Alamo2020b}. 
\subsection{Limitations of available data sources}
Although plenty of information is available for Covid-19 pandemic, it is also clear that important limitations are also presented in the available data sources (\cite[Section 4]{Alamo2020b}). The main limitations are: 
\begin{enumerate}
    \item Wide variety of data formats and structures, making it difficult to aggregate all the data in just one data set. 
    \item Time-varying nature of the sources, which limits simultaneous analysis on different locations.
    \item Some metrics do not reflect reality; for example, the number of confirmed cases underestimates the fraction of infected population. 
    \item Difficulty in calculating some characteristics of the virus: Because of the general lack of individual case data, relevant characteristics like latent and incubation periods, have to be inferred indirectly from aggregated time-series.  
    \item Variability on governments' criteria to make the data available. This translates in different formats and contents of the respective data-sets. Besides, many open sources are not properly documented. 
    \item Lack of transparency related to the real impact of the pandemic \cite{HoldenThorp2020}.
\end{enumerate}

These limitations undermine the use of the available raw data to i) measure the real state of the pandemic, ii) develop appropriated epidemic models, iii) assess the quality of governments' mitigation actions, and iv) plan ahead suitable strategies. 


Accurate models of Covid-19 pandemic cannot be developed just by using data related to the impact of the virus in human health. The reason is that the majority of parameters of the models cannot be explained isolating other related variables, such as demographic, connectivity, mobility,  and weather. Therefore, a wide variety of variables to enhance the prediction models are required. Similarly, these variables are also necessary to evaluate the impact of the assessment tools \cite[Section 3]{Alamo2020b}. 

\subsection{Covid-19 data sources}

Different open data sources with specific Covid-19 information are enumerated in this subsection. For a detailed list of resources, see \cite{alamo2020open}. 

\begin{itemize}
\item \textbf{Confirmed cases:}
Data sets collect the temporal series of the number of confirmed new cases, deaths, and recovered. Normally, data is available by country, and in some cases, also by region. The most used data set so far is maintained by Johns Hopkins University (JHU)\footnote{\url{https://github.com/CSSEGISandData/COVID-19}}. Other similar data sets can be found in \cite{alamo2020open}. Furthermore, local repositories for each country can also be found. On this line, Table \ref{tab:Regional Data} contains some example of regional data sets. This type of data sets presents a big picture of the pandemic in terms of human life impact.
\item \textbf{Pro-active testing:}
The data sets related to pro-active testing should provide information about the type of test carried out, the number of tests and number of positive cases. It is also essential to have access to auxiliary information such as age and gender group, professional activity, and the methodology used to carry out the selection of the individuals to be tested. The website \url{https://ourworldindata.org/} provides data about number of test carried out by country\footnote{\url{https://ourworldindata.org/coronavirus-testing}}.
\item \textbf{Contact-tracing:}
This data is related to contact among infected people and other persons. The data sets should indicate the connections of infected people with others in the last few days. In \cite{Ferretti2020}, the authors state that the speed of the spread of Covid-19 makes it impossible to implement manual tracing of contacts among infected people. Thus, it is clear that technology \cite{oliver2020mobile}, such as Internet of Thing (IoT) \cite{Klopfenstein2020} and wireless communications \cite{nanni2020:OLIVER:give}, should play an important role in this task. One important issue for measuring contacts and tracing citizens is data privacy. Currently, there is scarcity of data sets including contact-tracing (i.e., \cite{beraud2015french}). However, data sets used in multi-hop networks can be a direction to explore\footnote{\url{http://crawdad.org/}} \cite{khabbaz2011disruption} to develop models since this field has studied human behaviour in terms of contacts for two decades \cite{pelusi2006opportunistic}\cite{reina2015survey}. 
\item \textbf{Individual data:}
Individual data collection refers to data gathered directly from citizens. The individual data differs from the official data release from governments and institutions in many ways since it can be biased, and therefore, it should be analyzed carefully. Nevertheless, it is a useful source of data that should be taken into account to monitor the impact of individuals point of view. Moreover, pre-diagnosis can be done remotely using mobile devices, i.e., Apple has developed a pre-evaluation application\footnote{\url{https://www.apple.com/covid19/}}.        
\end{itemize}

\vspace{0.5cm}
 
\begin{table}[h]
\begin{tabular}{|
>{\columncolor[HTML]{DAE8FC}}c |
>{\columncolor[HTML]{67FD9A}}c |
>{\columncolor[HTML]{FFFC9E}}l |}
\hline
\multicolumn{1}{|l|}{\cellcolor[HTML]{FFCCC9}} & \cellcolor[HTML]{FFFC9E}\textbf{Source}                                                                        & \cellcolor[HTML]{ECF4FF} GitHub repositories \\ \hline
\textbf{Argentina}                             & 
\href{https://www.argentina.gob.ar/coronavirus/informe-diario}{Ministry of Health} &  
\href{https://github.com/SistemasMapache/Covid19arData}{Covid19arData}
\\ \hline
\textbf{Australia}                             & 
\href{https://www.health.gov.au/news/health-alerts/novel-coronavirus-2019-ncov-health-alert/coronavirus-covid-19-current-situation-and-case-numbers}{Australian Health Department} &    \href{https://github.com/covid-19-au/covid-19-au.github.io}{covid-19-au}       \\ \hline
\textbf{China}                                 & 
\href{http://en.nhc.gov.cn/DailyBriefing.html}{China National Health Commission} & 
\href{https://github.com/CSSEGISandData/COVID-19/}{JHU}, \href{https://github.com/midas-network/COVID-19/tree/master/data/cases/china}{Midas-China} \\ \hline
\textbf{France}                                & 
 \href{https://www.santepubliquefrance.fr}{Public France Health System} &
 \href{https://github.com/opencovid19-fr/data/blob/master/README.en.md}{opencovid19-fr}\\ \hline
\textbf{Germany}                               & 
\href{https://www.rki.de/EN/Home/homepage_node.html}{Robert Koch Institute} &     
\href{https://github.com/jgehrcke/covid-19-germany-gae}{covid-19-germany-gae}\\ \hline

\textbf{Iceland}                               & 
\href{https://www.covid.is/data}{Government of Iceland} & \href{https://github.com/gaui/covid19}{gaui-covid19}     
\\ \hline
\textbf{Italy}                                 & \href{http://www.protezionecivile.gov.it/attivita-rischi/rischio-sanitario/emergenze/coronavirus}{Italian Civil Protection Department}                                            &  
\href{https://github.com/pcm-dpc/COVID-19}{pcm-dpc} \\ \hline

\textbf{Paraguay}                              & 
 \href{https://www.mspbs.gov.py/reporte-covid19.html}{Ministry  of Public Health and Soc. Welfare} 
 &            
 \href{https://github.com/torresmateo/covidpy-rest/blob/master/data/covidpy.csv}{covidpy-rest}\\ \hline
\textbf{South Africa}                          &        
\href{https://www.nicd.ac.za/}{National Inst. Communicable Diseases}&  \href{https://github.com/dsfsi/covid19za}{covid19za}                               \\ \hline
\textbf{South Korea}                          &   
\href{https://www.cdc.go.kr/board/board.es?mid=a30402000000&bid=0030}{Centers for Disease Control and Prevention}&    
 \href{https://github.com/parksw3/COVID19-Korea}{COVID19-Korea} \\ \hline
\textbf{Spain}                          &   
\href{https://www.mscbs.gob.es/profesionales/saludPublica/ccayes/alertasActual/nCov-China/home.htm}{Ministry of Health}&    
\href{https://github.com/datadista/datasets/tree/master/COVID\%2019}{datadista-Covid-19} \\ \hline
 \textbf{United Kingdom}                        &    
 \href{https://www.gov.uk/government/publications/covid-19-track-coronavirus-cases}{Pubic Health England} &                 \href{https://github.com/tomwhite/covid-19-uk-data}{covid-19-uk-data}                 \\ \hline
\textbf{United States}                         &  
 \href{https://www.cdc.gov/}{Centers for Disease Control and Prevention} &   
  \href{https://github.com/CSSEGISandData/COVID-19/}{JHU}, \href{https://github.com/nytimes/covid-19-data}{Nytimes}\\ \hline
\end{tabular}
\caption{\label{tab:Regional Data} Some examples of regional Covid-19 data resources. See more examples of open data sets in \cite{Alamo2020b}. }
\end{table}

\subsection{Government measures}
 The level and severity of the executed strategies are variable, ranging from soft actions, like encouraging social distance and mask use, to hard measures such as closing schools, forbid massive events, and complete lock-down. In this context, in \cite{hale2020variation}, the authors developed the Stringency Index (SI)\footnote{\url{https://www.bsg.ox.ac.uk/research/research-projects/coronavirus-government-response-tracker}}, that captures variation in containment and closure policies. SI considers 17 indicators of government responses, including containment and closure policies,  economic policies, and health system policies. The values of SI are within the interval $[0, 100]$, being a value of 100 the strictest response. 

\subsection{Social-economic indicators}

Different data resources on social-economic indicators are presented in what follows. See also \cite{alamo2020open}. 
%

\begin{itemize}
    \item \textbf{Mobility}: Data regarding to mobility refers to reports on changes of mobility patterns \cite{warren2020mobility}. For instance, community mobility. On this line, both Google\footnote{\url{https://www.google.com/covid19/mobility/}} and Apple\footnote{\url{https://www.apple.com/covid19/mobility}} present mobility reports by location.
    \item \textbf{Online questionnaires}: There are several data sources that collected individual data regarding the social impact of Covid-19 in citizens, such as \cite{oliver2020covid19impact} \cite{qiu2020nationwide} and \cite{kleinberg2020measuring}. 
    \item \textbf{Social networks}: Several social networks like Facebook and Twitter\footnote{\url{https://ieee-dataport.org/open-access/corona-virus-covid-19-geolocation-based-sentiment-data}} \cite{shanthakumar2020understanding} allow users to post their emotion and feelings \cite{kleinberg2020measuring}.
    \item \textbf{Internet search}: Google searches and Baidu Search Index (BSI) are good examples of social indicators. Although not longer available, Google Flue application was created to measure the trends of Google queries about flu and dengue. The historical data is still available for analysis\footnote{\url{https://www.google.org/flutrends/about/}}.
 
\end{itemize}

\subsection{Auxilary data sources}

In this subsection, we include datasets relevant for the study and development of models of Covid-19, such as
health-care system, demography, weather and air transport connectivity. These are variables that are under research
to evaluate their influence on virus propagation. See \cite{alamo2020open} for a more detailed enumeration. 

\begin{itemize}
    \item \textbf{Health-care system}:
Data related to health care systems should include, among others, the number of Intensive Care Unit (ICU) and ventilators. The data about health-care resources is maintained by the national health-care systems of each country, and only partially accessible in some data-sets.
\item \textbf{Demographics}:
The normalization of the Covid-19 data should be carried out considering demographic data to develop general models and actions\footnote{\url{https://www.kaggle.com/tanuprabhu/population-by-country-2020}}. Population density is also relevant to explain the rapid spread of the virus in some locations. Besides, age groups are relevant to infer the mortality of Covid-19\footnote{\url{https://ourworldindata.org/age-structure}}.  
\item \textbf{Weather and climate data}:
The seasonal behavior of Covid-19 is under study and discussion \cite{shi2020impact} \cite{WU2020139051} \cite{sajadi2020temperature}. Data sets on weather and climate should include variables that affect the spread of the virus. For instance, high temperature and humidity reduce the spread of the virus \cite{otter2016transmission}. There are several institutions that provide weather data \cite{alamo2020open}, such as the European Centre for Medium-Range Weather Forecasts (ECMWF) and the National Oceanic and Atmospheric Administration (NOAA). Regarding climate change, some reports indicate that pollution levels also favors the spread of the virus \cite{ZHU2020138704}\cite{Prathereabc6197}. 
\item \textbf{Air transport connectivity}: International air routes explain the propagation of Covid-19 from Wuhan outbreak to other territories \cite{gilbert2020preparedness} \cite{Haider2020}. Datasets on air transport connectivity should contain information on passengers and routes\footnote{\url{https://flirt.eha.io/}}. 
\end{itemize}

\subsection{Data curation}

Data curation is the active management of data over its life cycle to ensure it
meets the necessary data quality requirements for its effective usage \cite{lord2004data}.
Data curation processes can be categorised into different activities such as content creation, selection, classification, transformation, validation, and preservation \cite{freitas2016big:data:curation}. 

Covid-19 data-driven methods require that data is trustworthy, discoverable, accessible, reusable, and frequently updated. 
A key trend for the curation of Covid-19 data sets are different open-source communities like Kaggle and GitHub \cite{alamo2020open}.

\section{Consolidated Time-Series}
\label{Sec:Consolidated:Time:Series}
The available data collections and resources offer many opportunities in the context of monitoring, modelling and decision-making. However, this is intrinsically tied to the trust we can put in the origins and quality of the underlying data \cite{Sanger2014}.

In this section, We review some of the most relevant theories and methodologies that can be used to process raw data from different and heterogeneous sources in order to obtain consolidated time-series serving to monitor Covid-19 pandemic. The goals are i) detect and correct inconsistencies in the raw data; ii) take advantage of spatial and time similarities in segregated regional data to produce enhanced aggregated time series; iii) detect regional clusters with similar characteristics; iv) statistically characterize the interplay between different variables  in order to provided enhanced estimations of the real dynamics of the pandemic. 

The methodologies presented in this section have not clear borders distinguishing one from each other because they share many tools and approaches.
The nomenclature can vary across the different disciplines using these methodologies (data science, epidemiology, control theory, to name just a few). 

\subsection{Pre-processing data}

In the field of epidemiology,  it is usual to employ the term ``cleaning data" to refer to the process in which one identifies the errors in collected data and corrects them, or at least minimizes their effects on subsequent analysis. As detailed in  \cite{VandenBroeck2013}, this process involves three steps: screening of the data, detecting possible outliers, and editing data abnormalities. Another standard procedure is the initial normalization of the data, ensuring that each of the values is included in the $[0,1]$ interval. This translates, for example, in dividing by the total size of the population, the counts of confirmed and death cases. This is relevant when comparing the impact of the epidemic in different regions. Besides, another important procedure is standardization, in which the original value $x$ of a given variable is replaced by the quotient $(x-\mu)/\sigma$, where $\mu$ and $\sigma$ are the empirical mean and standard deviation of the variable under consideration. The standardization process can be applied in auxiliary variables like temperature and humidity. Normalization and standardization are common procedures in clustering methods because they facilitate the comparison between variables and the computation of dissimilarity indices \cite{Trebuna2014:Normalization}, \cite[Chapter 2]{Wierzchon2018:CLUSTERING}.

\subsection{Data reconciliation }

Data reconciliation is a methodology that incorporates prior knowledge on the data to enhance its consistency (see e.g. \cite{Albuquerque1996}, \cite{narasimhan1999data}). The idea is to ``reconcile" the data with some initial assumptions. Suppose, for example, that $x\in \R^n$ is known to satisfy the constraints $Ax=b$ and $Cx\preceq d$, where 
$\preceq$ denotes componentwise inequalities. If we have an estimation $x^e$ on $x$ that does not satisfy the constraints, then the reconciled value $x^r$ for $x$ is obtained from $x^r=x^e+\Delta x$, where $\Delta x$ is obtained from the solution of the following optimization problem
\begin{eqnarray*}
\min_{\Delta x} && \| \Delta x \|^2\\
\mbox{s.t.} &&  A(x^e+\Delta x)= b\\
&& C(x^e+\Delta x)\preceq d
\end{eqnarray*}
where $\| \cdot \|$ denotes a possibly weighted Euclidean norm. The data reconciliation procedure is written as a projection problem: computing the minimum distance to a convex set. In the presence of only equality constraints, the reconciliation optimization problem can be solved by means of the solution of a linear system of equations. From the theory of projections operators \cite{Deutsch2012}, \cite[Lemma 2.2.8]{Nesterov2018} we have that if $x^e$ is not consistent with the assumptions $Ax=b$ and $Cx\preceq d$, then the reconciled value $x^r=x^e+\Delta x$ is closer to $x$ than the initial value $x^e$. More specifically,
$$ \| x^r- x\|^2\leq \| x^e - x\|^2- \|\Delta x\|^2.$$

{\bf Illustrative example:}  Suppose that $(I_a, D_a)$ and $(I_b,D_b)$ represent the accumulated counts of confirmed and death cases in regions $R_a$ and $R_b$ respectively. If we assume that the probability of dying, once infected, is similar in both regions the quantities $\frac{D_a}{I_a}$ and $\frac{D_b}{I_b}$ should be similar (equivalently $D_aI_b$ should be close to $D_bI_a$). If one is confident about the accuracy on the number of death counts $D_a$ and $D_b$, but not too much on the values $I_a$ and $I_b$, one could reconcile them using the assumption that the probability of dying is equal in both regions. The theory of data reconciliation states that the reconciled values for $I_a^r$ and $I_b^r$ are 
$I_a^r=I_a+\Delta I_a$ and $I_b^r=I_b+\Delta I_b$ where $\Delta I_a$ and $\Delta I_b$ are obtained from the solution of the following optimization problem.
\begin{eqnarray*}
\min_{\Delta I_a, \Delta I_b} && \left(\frac{\Delta I_a}{N_a}\right)^2 + \left(\frac{\Delta I_b}{N_b}\right)^2\\
\mbox{s.t.} &&  (I_a+\Delta I_a)D_b = (I_b+\Delta I_b)D_a,
\end{eqnarray*}
where $N_a$ and $N_b$ are the sizes of the populations of regions $R_a$ and $R_b$. We notice that the optimization problem obtains the minimal norm modification on the fraction of confirmed cases that forces the assumption on equal death risk to be satisfied. 

As commented in Section \ref{sec:data:sources}, there are many limitations and inaccuracies in the time-series related to Covid-19. On the other hand, there are multiple data sets corresponding to different locations. Identifying valid equality and inequality constraints for the time series provides, along with the data reconciliation methodology, a powerful tool to obtain consolidated time-series. In this way, inaccuracies in the available data are significantly compensated. 
 
 \subsection{Data fusion}
 
 Data fusion is defined as the theory, techniques and tools that are used to combine data from different sources into a common representation format \cite{Mitchell2012}, \cite{Nguyen2012}. Simultaneous monitoring of signals coming from different sources, or surveying diverse aspects of data, even if it comes from a single source, can yield improvement in accuracy over more traditional univariate analyses \cite{Dubrawski2011}.
 
 Data fusion is very valuable in the context of Covid-19 because, in many situations, a single variable, or single time-series has to be obtained from numerous data sources (for example, the evolution of a single mobility index is inferred from the geo-localization of a large number of individuals). Specific techniques like temporal alignment \cite[Chapter 6]{Mitchell2012} can also be used to compare better two time-series corresponding to epidemic data of different locations.  
 
 \subsection{Clustering methods}
 
 Cluster analysis restructures the data into clusters sharing common statistical characteristics \cite{Wierzchon2018:CLUSTERING}, \cite{BrianS.EverittSabineLandauMorvenLeese2011}. 
 Seasonality, climate, geography, measurement strategies or implemented mitigating measures are possible reasons for the existence of cluster formation. Being able to detect such configuration is important prior to further analysis. For example, spatial clustering is used in \cite{Chan2011:cLUSTERING:SURVEILLANCE} to better assess the thresholds required to detect the outbreak of an infectious disease. In \cite{gilbert2020preparedness}, clustering is used to group countries with
similar importation patterns of Covid-19 cases in African countries.
 
One is advised to apply a clustering method prior to further analysis because the existence of clusters may
impact the results, as intra-cluster correlation may exist so that the frequent independence assumptions are violated \cite{Wierzchon2018:CLUSTERING}. Moreover, cluster analysis allows to assess better why a specific control strategy is working better in one location than in another. Actions applied in locations belonging to the same cluster have a larger probability of yielding similar outcomes.

 Cluster analysis in a spatial setting has already been used in the monitoring and understanding of SARS epidemic \cite{Lai2004:clustER}, \cite{Small2005}. The number of detected clusters and their spatial distribution is relevant to the effectiveness of control measures \cite{Small2005}. 
 
 \subsection{Time-series theory}

In this subsection, some mature techniques from signal processing \cite{Oppenheim2015}, time series analysis \cite{Durbin2012} and stochastic processes \cite{Papoulis2002,Stark2012:Signal:Processing} are commented. They can be applied to enhance the quality of the raw time-series and these are used, for example, to characterize the raw daily prevalence data of Covid-19 \cite{Benvenuto2020}. 

\begin{itemize}
    \item {\bf Discrete signal processing:} Filtering methods, based on the characterization of the signals on the frequency domain, are used to smooth signals reducing the effect of high frequency noisy signals due to measurement errors \cite{Papoulis2002}. Moving averaging filters are used, for example, to smooth the signal corresponding to the daily death counts \cite[Subsection 7.1.1]{Alamo2020b}.
    \item {\bf State space methods:} State space modelling provides a unified methodology for treating a wide range of problems in time series analysis: filtering, smoothing, and forecasting. In this approach, it is assumed that the development over time of the system under study is determined by an unobserved series of variables associated to a series of observations. The relationship between observed and unobserved variables is done by means of a state space representation \cite{Durbin2012}. A key ingredient of these methodologies is the Kalman Filter, which is employed to obtain an estimation of the non-observed variables. A direct application is the smoothing and filtering of the signal: one can consider the noisy signal as the observed one and the noise-free signal as the unobserved. This leads to the implementation of methods to enhance the quality of the processed signals \cite[Chapter 9]{Stark2012:Signal:Processing}.
    \item {\bf Stochastic processes:} The theory of stochastic processes \cite{Papoulis2002} allows us to characterize in a probabilistic way the time series related to the epidemic. This characterization is the base of statistical signal processing (see \cite[Chapter 9]{Stark2012:Signal:Processing}) that is used to estimate random variables and implement Kalman filtering approaches. 
\end{itemize}

All these techniques can find a direct application in the context of the consolidation of the historical data of an epidemic. Besides, statistic signal processing is one of the pillars of different inference methods required to fit epidemiological models to data (see Subsection \ref{subsection:Fitting:to:Data}).


\section{Estimation of the state of the pandemic}\label{sec:Estimation:State:Epidemic}

The control of Covid-19 pandemic requires monitoring of essential indicators. This includes not only estimations of the current incidence of the disease in the population, but also the (daily) surveillance of measures such as social-distancing, mobility and others, with direct effects on its spread. Monitoring is key in the decision-making process because it provides fundamental information to decide whether to lift or strengthen measures and restrictions. 

A pandemic outbreak, or a recurring wave, needs an immediate response, which requires up-to-date estimations of the state of the pandemic. This estimation process is hindered by the incubation period of Covid-19, which introduces a time-delay between the infection and its potential detection. Another issue is the infectious asymptomatic population, which is an important transmission vector difficult to detect \cite{Ferretti2020}. All these shortcomings motivate the deployment of specific surveillance and estimation methodologies capable of using the available information to enable quick adjustment of different control measures.

In this section, we cover the most relevant techniques to monitor the state of the pandemic, focusing on the approaches that are oriented to i) real-time monitoring of different aspects of the pandemic (real-time epidemiology), ii) early detection of infected cases and immune response estimation  (pro-active testing);  iii) estimation of the current fraction of infected population, symptomatic or not (state estimation methods); v) early detection of new waves (epidemic wave surveillance).

\subsection{Real-time epidemiology} 
The use of a large number of real-time data streams to infer the status and dynamics of the public health of a population presents enormous opportunities as well as significant scientific and technological challenges \cite{Bettencourt2007}, \cite{Zeng2010:BioSurveillance}, \cite{Drew2020:science:real:time:epidemiology}. 

Real-time Covid-19 data can be of very different nature and origin (e.g. mobile phone data, social media data, IoT data and public health systems) \cite{ting2020digital:data}. Mobile phone data, when used properly and carefully, represents a critical arsenal of tools for supporting public health actions across early, middle, and late-stage phases of the Covid-19 pandemic \cite{Oliver2020:MOBILE:DATA:SCIENCE}. 
Voluntary installed Covid-19 apps or web-based tools enable self-report of data related to exposure and infections. The information steaming from these sources provide real-time scalable epidemiological data, which is used to identify populations with highly prevalent symptoms that may emerge as hot spots for outbreaks \cite{Drew2020:science:real:time:epidemiology}. In this context, it is also important to mention social media, which is relevant to assess the mobility of the population and its awareness with regard to social distancing, the state of the economy and many other key indicators \cite{Zhou2020GIS:BIG:DATA}, \cite{cinelli2020covid}. 

The magnitude and scale of population mobility are essential information for spatial transmission prediction, risk area division, and control measure decision-making for infectious diseases. Nowadays, the most effective tool to access this sort of real-time information is through Big Data technologies and Geographic Information Systems (GIS). These systems have played a relevant role when addressing past epidemics like SARS and MERS \cite{Peeri2020:SARS:MERS:IoT}, providing efficient aggregation of multi-source big data, rapid visualization of epidemic information, spatial tracking of confirmed cases, surveillance of regional transmission and spatial segmentation of the epidemic risk \cite{Zhou2020GIS:BIG:DATA}, \cite{Wang2020BIGDATA}. 

\subsection{Proactive testing}
Proactive testing is key in the control of infectious diseases because it provides a way to identify and isolate the infected population. Besides, it is also a relevant source of information to identify risk areas, percentage of asymptomatic carriers and attained levels of immunology response in the population \cite{winter2020important:SEROLOGY}.   

There are different methodologies to approach proactive testing:
\begin{itemize}
    \item \textbf{Risk based approach}: The individuals with the highest probability of being carriers of the disease are tested. This implies testing not only the individuals with symptoms but also the ones that are more exposed. For example, health-care workers are at high risk and can also be relevant vectors (this is also the case in other professional sectors). In a second level, we have the individuals that thanks to personal contact-tracing have been identified to be more exposed to a specific confirmed case. In a third level, we find the individuals that have travelled, of often go, to hot spots of the pandemic \cite{Wang2020BIGDATA}. The determination of risk zones can be done by means of government mobility surveillance or by personal software environments \cite{Drew2020:science:real:time:epidemiology}.
   \item \textbf{Voucher-based system}: People who test positive are given an anonymous voucher which they can share with a limited number of people whom they think they might be infected. The recipients can use this voucher to book a Covid-19 test and can receive their test results without ever revealing their identity. People receiving positive result are given vouchers to further backtrack the path of infection \cite{roomp:Oliver:2020acdc}. See also \cite{nanni2020:OLIVER:give}. 
   \item \textbf{Serology studies}: One of the main limitations of  RT-PCR tests is its inability to detect past infection. Serological testing carried out within the correct time frame after disease onset can detect both active and past infections. Furthermore, serological analysis can be useful to define clusters of cases, retrospectively delineate transmission chains and ascertain how long transmission has been ongoing or to estimate the proportion of asymptomatic individuals in the population \cite{winter2020important:SEROLOGY}.
   
   \end{itemize}


\subsection{State space estimation methods} 

Dynamic state-space epidemiological models are fundamental to characterize how the virus spreads in a specific region and to estimate not directly measurable time-varying epidemiological variables \cite{Cazelles1997}. Classical state space estimation methods like Kalman filter \cite{narci:hal-02475936}, \cite{Riad2019} are employed to estimate the fraction of current infected population. The objective of Kalman filter is to update knowledge about the state of the system each time a new observation is available \cite{Durbin2012}. Different modifications and generalizations of Kalman filter are also able to address the specifics of an epidemic model. These methodologies are essential both to the estimation problem and to the inference of the parameters that describe the model (see \cite{Schon2011} and \cite{Abreu2020}). 

\subsection{Epidemic wave surveillance}

Based on current evidence, the most plausible scenario may involve recurring epidemic waves interspersed with periods of low-level transmission \cite{WHO:PHSM:lift:2020}. In this context, it is crucial to implement a surveillance system able to detect or anticipate, possible recurring epidemic waves. These systems enable an immediate response that reduces the potential burden of the outbreak.

Outbreak detection relies on methodologies able to process a large amount of data steaming from the different surveillance systems \cite{DelValle2020:Secondary:Wave}, \cite{Althouse2015:SURVEILLANCE:NOVEL:DATA:STREAMS}. With this information, mechanisms to determine if the spread of the virus has surpassed a threshold requiring mitigation measures can be implemented, see, e.g. \cite{Lazarus2010:Surveillance}. There is a large body of literature on this epidemiological detection problem since many infectious diseases undergo considerable seasonal fluctuations with peaks seriously impacting the health-care systems \cite{Sparks2013}, \cite{Unkel2012:REVIEW:sURVEILLACE}. National surveillance systems are implemented worldwide to detect influenza-like illnesses outbreaks rapidly, and assess the effectiveness of influenza vaccines \cite{Vega2015}, \cite{ECDC:2013:SEASONAL:INFLUENZA}. Specific methodologies to determine the baseline influenza activity and epidemic thresholds have been proposed and implemented \cite{Vega2013:MOVING:EPIDEMIC:METHOD}. The focus of these methods is to reduce false alerts and detection lags. Outbreak detection can be implemented in different ways that range from simple predictors based on moving average filters  \cite{Farrington1996} and fusion methods \cite{Dubrawski2011} to complex spatial and temporal clustering \cite{Chan2011:cLUSTERING:SURVEILLANCE}. 

The detection of Covid-19 recurring epidemic waves poses new challenges for several reasons: i) lack of historical seasonal data, ii) difficulties in determining the current fraction of infected population, and iii) computation of baselines and thresholds demands a precise characterization of the regional (time-varying) reproduction number.



\section{Epidemiological models} \label{sec:models}

Epidemiology is a well established field \cite{Martcheva2015} that models the spread of infectious diseases. Given the high complexity of these phenomena, models are key to synthesize information to understand epidemiological patterns and support decision making processes \cite{Heesterbeek2015}.

\subsection{Time-response and viral shedding of Covid-19}

The available epidemiological and clinical studies of the virus allow to model it from a time evolution point of view \cite{Li2020:virus}, \cite{wang2020unique}, \cite{Hellewell2020}. How the disease and its potential infectious evolves with time is characterized by means of the following key epidemiological parameters (see e.g. \cite{he2020temporal} and \cite{Hellewell2020}):

\begin{itemize}
    \item {\bf Latent time}: time during which an individual is infected but not yet infectious. Initial estimates are of 3-4 days \cite{Li2020:virus}.
    \item {\bf Incubation time}: the time between infection and onset of symptoms. The median incubation period is estimated to be 5.1 days, and 97.5\% of those who develop symptoms will do so within 11.5 days of infection \cite{Lauer2020:INCUBATION}. The median time between the onset of symptoms to death is close to 3 weeks \cite{Zhou2020:CLINICAL:COURSE}.
    \item {\bf Serial interval}: time step between symptom onsets of successive cases in a transmission chain.
    Initial estimates of the median serial interval yield a value of around 4 days, which is shorter than its median incubation period \cite{Nishiura2020:Serial:Interval}. This implies that a substantial proportion of secondary transmission may occur prior to illness onset.
    \item {\bf Infectiousness profile}: characterizes the infectiousness of an infected individual along time. In \cite{Zhou2020:CLINICAL:COURSE}, the median duration of viral shedding estimation was 20 days in survivors while the most prolonged observed duration of viral shedding in survivors was 37 days.
    \item {\bf Basic reproduction number $R_{0}$}: represents the average number of new infections generated by an infectious person at the early stages of the outbreak, when everyone is susceptible, and no countermeasures have been taken \cite{liu2020reproductive}, \cite{Park2020:REproductive:Number}. First estimations range from 2.24 to 3.58 \cite{Zhao2020}. The effect of temperature and humidity in this parameter is addressed in different studies. See, for example \cite{Mecenas2020:TEMPERATURE} and Subsection \ref{sub:sub:sec:Seasonal}.
\end{itemize}

   The basic reproduction number, along with the serial interval, can be used to estimate the number of infections that are caused by a single case in a given time period. Without any control measure, at the early stages of the outbreak, more than 400 people can be infected by one single Covid-19 case in one month \cite{Nicola2020:REVIEW:Management}. Estimates of the basic reproductive number are of interest during an outbreak because they provide information about the level of intervention required to interrupt transmission and about the potential final size of the outbreak \cite{Park2020:REproductive:Number}.
     
We notice that the aforementioned parameters are often inferred from
epidemiological models, once they have been fitted to the available data on the number of confirmed cases and dying patients.

\subsection{Compartmental models}
The idea underneath a compartmental model is dividing a population into different groups or compartments. Each compartment tracks the number of individuals in the same state of the epidemic.

\subsubsection{Basic compartmental models}\label{sec:basic:compartmental:models}

The simplest compartmental approach is the so-called SIR model, introduced by Kermack and McKendrick at the beginning of the 20th century. The model has only three compartments: \emph{Susceptible} (S), representing healthy individuals susceptible of getting infected, \emph{Infected} (R), and \emph{Recovered} (R). This last compartment can also take into account deceased individuals. Nevertheless, for low mortality rate diseases,  including only recovered individuals, is a good approximation. 

The dynamics of an epidemic using a SIR model can be written as follows:
\begin{eqnarray}
\frac{dS(t)}{dt} &=& -\beta S(t)\frac{I(t)}{N}, \label{eq:SIR_S} \\
\frac{dI(t)}{dt} &=& +\beta S(t)\frac{I(t)}{N} - \mu I(t), \label{eq:SIR_I}\\
\frac{dR(t)}{dt} &=& \mu I(t),\label{eq:SIR_R}
\end{eqnarray}

where $N$ represents the total population size, $\beta$ is the rate of infection, and $\mu$ is the recovery rate.
At the beginning of an epidemic $S$ equals approximately the entire population, and thus from \eqref{eq:SIR_I} it holds that $I(t) = I_{0}e^{(\beta-\mu)t} =  I_{0}e^{\mu(R_{0}-1)t}$, where $I_0$ represents the initial number of infected $I_0=I(0)$ and $R_{0} = \beta / \mu$ is the \emph{basic reproduction number} mentioned in the previous section. This number can be understood as the average number of secondary cases produced by an infectious individual. Clearly, when $R_{0}$ is greater than 1, there is an exponential increase in the number of infected individuals on the early days of the epidemic. The same equation can also be used to estimate the point at which the rate of newly infected individuals begins to fall $S(t)<N/R_{0}$. At this point, the given population has reached what is known as \emph{herd immunity}.

To account for the incubation time, an enhanced version of SIR model, the SEIR model, includes an extra compartment: \emph{Exposed} (E). Exposed individuals are not able to transmit the disease yet, but are transferred to the Infectious compartment with a fixed rate, modelling the incubation period.

SEIR models have been recently used to analyze Covid-19 pandemic. For example, in \cite{Fang2020} and \cite{Kucharski2020}, the spread dynamics and different control measures are modelled with a SEIR model. The parameters of the model are adjusted by simulation and data fitting. A SEIR model is also used in \cite{Wu2020} fitted with data from Wuhan. This model is improved in \cite{Wu2020b} to estimate clinical severity.  
\vspace{2mm}

\subsubsection{Extended compartmental models}

\begin{figure}[ht!]
    \centering
    \includegraphics[width=0.8\textwidth]{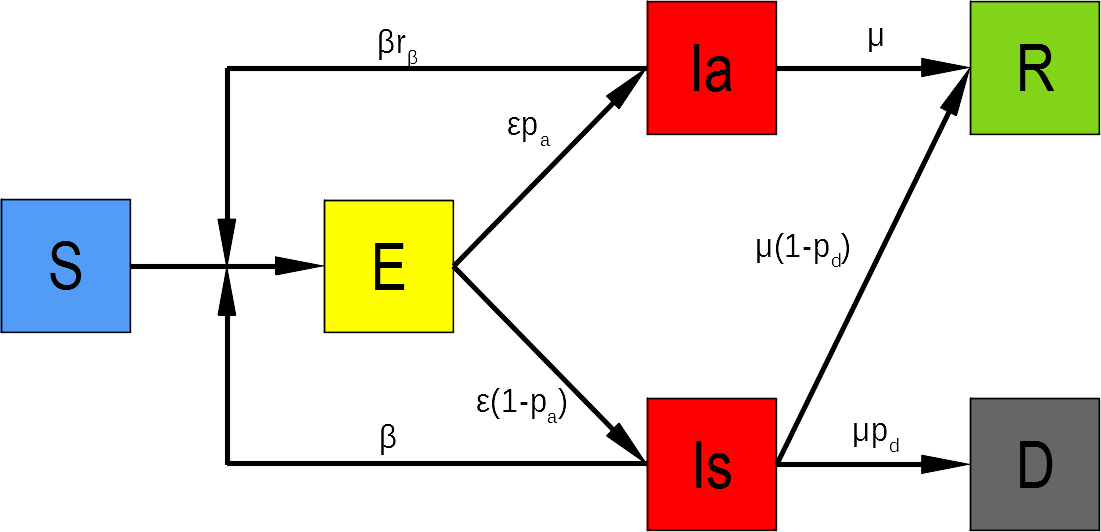}
    \caption{Illustration of an extended compartment epidemic model with six compartments: Susceptible (S), Expose (E), Asymptomatic Infected (Ia), Symptomatic Infected (Is), Recovered (R), and Dead (D). The $\beta$ value is the transmission rate, $\epsilon$ is the expose rate and $\mu$ is the recovery rate (including deaths). The parameter $r_{\beta}$ determines the contribution of asymptomatic infected individuals to the transmission rate. The term $p_a$ represents the probability that an infected individual becomes an asymptomatic one. The term $p_d$ reflects the probability of death of an symptomatic infected individual}
    \label{fig:epidemic_model}
\end{figure}

Further extended versions of compartmental models include extra compartments and transitions between them, as for instance, symptomatic and asymptomatic individuals (see Figure \ref{fig:epidemic_model}), the possibility of re-infection after recovery, or individuals in quarantine \cite{Crokidakis2020}.

Many applications of these extended models can be found in the literature. For example, in \cite{Riley2003}, the authors used a dynamical compartmental model to analyze the effective transmission rate of SARS epidemic in Hong Kong. The model consisted of 7 compartments: susceptible, latent, infectious, hospitalized, recovered, and dead individuals. Moreover, in \cite{Chowell2014}, a stochastic SEIR model is used to estimate the basic reproduction number of MERS-CoV in the Arabian Peninsula. The compartments distinguish between cases transmitted by animals and secondary cases, and the estimation of the model parameters employs a delayed rejection adaptive Metropolis-Hastings algorithm in a Markov-Chain Montecarlo framework.

In the case of Covid-19 pandemic,  asymptomatic infected people play an important role in the spread of the Covid-19 (see  \cite{Giordano2020} and \cite{Ferretti2020}). In \cite{Giordano2020}, a SIDARTHE model is proposed. The total population is partitioned into: S, Susceptible; I, Infected (asymptomatic infected, undetected); D, Diagnosed (asymptomatic infected, detected); A, Ailing (symptomatic infected, undetected); R, Recognised (symptomatic infected, detected); T, Threatened (infected with life-threatening symptoms, detected); H, Healed (recovered); E, Extinct (dead). The interactions among these eight stages are modelled by a set of parameters. In \cite{Ferretti2020}, the epidemic model considers a transmission rate $\beta$ that takes into account the contributions of asymptomatic transmission, presymptomatic (asymptomatic) transmission, symptomatic transmission, and environmental transmission compartments. The results indicate that the contribution of asymptomatic infected to $R_{0}$ is higher than symptomatic infected and other ways of transmission. The main reason is that symptomatic infected are often rapidly detected and isolated.

\subsubsection{Age-structured models} Age-structured epidemic models make it possible to relax random mixing hypothesis incorporating heterogeneous, age-dependent contact rates between individuals \cite{Valle2013:age:GROUPS}. In \cite{Xue-Zhi2001} and \cite{Safi2013}, stability results for different age-structured SEIR models are given. For Covid-19, an age-structure model, aiming at estimating the effect of social distancing measures in Wuhan, is presented in  \cite{Prem2020}. In \cite{Saljeeabc3517}, a stratified approach is used to model the epidemic in France. 

\subsubsection{Modelling the seasonal behaviour of Covid-19}\label{sub:sub:sec:Seasonal}
Some works have studied the influence on temperature and humidity in the spread of Covid-19 (e.g. \cite{Mecenas2020:TEMPERATURE} and \cite{sajadi2020temperature}). It has been reported that both variables have an effect on the basic reproduction number $R_{0}$ \cite{wang3551767high}. As an example, the results in \cite{wang3551767high} indicate that an increment of one-degree Celsius in temperature and one per cent in relative humidity lower $R_{0}$ by 0.0225 and 0.0158.  This influence should be included in the epidemic models to capture the seasonal behaviour of Covid-19.  For instance, by considering the parameters $\beta$ and $\mu$ functions of both temperature \cite{waikhom2016sensitivity} and relative humidity. Yet it remains unclear whether seasonal and geographic variations in climate can substantially alter the pandemic trajectory, given high susceptibility is a core driver \cite{Baker2020:Seasonal:Limit}.

\subsection{Spatial epidemilogy}
One of the flaws of compartmental models is that they were developed to describe the evolution of epidemics in a single population where each individual is assumed to interact with every other at a common rate (homogeneous contact).  This can be a reasonable approximation within a given population, but it is not appropriate to study the global spread of a pandemic.

In the last decade, compartmental models have been extended successfully to spatial epidemiology models in order to analyze spreading phenomena in a more accurate way.

\begin{itemize}
    \item \textbf{Metapopulation models}: Metapopulation models integrate two types of dynamics: the one related to the disease, typically driven by a compartmental model, and the mobility of individuals (agent-based model) across the subpopulations that build the metapopulation under analysis. As a representative example, in \cite{Brockmann2013} the authors introduce the notion of effective distance to capture the spatio-temporal dynamics of epidemics, combining the SIR model of $n=1,2,\ldots ,p$ populations with mobility among them. The resulting model for each population is 

\begin{eqnarray}
\frac{dS_{n}(t)}{dt} &=& -\beta S_{n}(t)\frac{I(t)}{N_{n}} + \sum_{m\neq n}{(w_{nm}S_{m} - w_{mn}S_{n})}, \label{eq:meta_SIR_S} \\
\frac{dI_{n}(t)}{dt} &=& +\beta S_{n}(t)\frac{I_{n}(t)}{N_{n}} - \mu I_{n}(t) + \sum_{m\neq n}{(w_{nm}I_{m} - w_{mn}I_{n})}, \label{eq:meta_SIR_I}\\
\frac{dR_{n}(t)}{dt} &=& \mu I_{n}(t) + \sum_{m\neq n}{(w_{nm}R_{m} - R_{mn}I_{n})}.\label{eq:meta_SIR_R}
\end{eqnarray}

In this model, $w_{nm}$ is the per capita traffic flux from population $n$ to population $m$, given by $w_{nm}=F_{nm}/N_{m}$, where $F_{nm}$ is the total flux and $N_{m}$ is the size of the population $m$. In \cite{Aleta}, the authors used a SEIR compartmental model together with stochastic data-driven simulations to capture the mobility in all Spanish provinces. The work focuses on evaluating the effectiveness of contention measurements in Spain on February 28th, when a few dozen cases of Covid-19 had been detected. By capturing both temporal and spatial evolution of epidemics, metapopulation models are also capable of forecasting the effectiveness of mobility restrictions.
\item \textbf{Social networks models}: Another approach to address the same problem is based on social networks models \cite{El-Sayed2012}. These models consider that transmission can only occur along linked or connected individuals, which makes it possible to model heterogeneity in contact patterns in an explicit manner. Small-world networks have been used in combination with compartmental models to model disease transmission of SARSs \cite{Small2005} and Covid-19 \cite{Brethouwer2020}, and also to assess the efficacy of measurements as contact tracing \cite{Kiss2006}. In general, network models produce a more accurate prediction of the disease spread.  In particular, the use of homogeneous compartmental models in population with heterogeneous contacts tends to underestimate disease burden early in the outbreak and overestimate it towards the end, although for certain kind of networks it is possible to modify compartmental models to fix this problem \cite{Bansal2007}. Another interesting aspect of studying epidemic spread with network models is the observation of the percolation phase transition. That is, a change on the connection among nodes that abruptly modifies the global connectivity of a graph. Percolation theory has been widely studied in random networks
\cite{albert2002statistical}. In the context of epidemic modelling, the transition phase occurs where isolated clusters of infected people join to form a giant component that is able to infect many people \cite{harding2020phase}.     

\end{itemize}
\subsection{Computer-based models}

Computer-based simulation methods to predict the spread of epidemics can take into account numerous factors, such as heterogeneous behavioural patterns, mobility patterns, both at long and short scales, demographics, epidemiological data, or disease-specific mechanisms \cite{Helbing2015}.

As a representative example, the Global Epidemic and Mobility simulation framework (GLEAM) allows performing stochastic simulations of a global epidemic with different global-local mobility patterns, as well as data regarding demographics or hospitalization \cite{VanDenBroeck2011}.

However, detailed simulation-based methods depend on a significant number of parameters, which need to be chosen a fixed for a specific simulation. This is especially difficult in the early days of an epidemic outbreak. Furthermore, these approaches do not reveal which factors are actually relevant in the spread of epidemics. 

Simpler data-driven tools have also been developed to overcome these difficulties. For example, in \cite{Huang2020}, a model-free tool based on daily newly confirmed and recovery cases is developed. As no model is used to calculate how asymptomatic infected infect others, the method does not need information about infection rate, asymptomatic infected or susceptible individuals. The disadvantage is that the method cannot be used when the epidemic has started, and the data is incomplete.

\subsection{Modelling the effect of containment measures}

Controlling an emerging communicable disease requires both the prompt implementation of measures and the rapid assessment of their efficacy \cite{Cauchemez2006:Assessment:efficacy}. In what follows, we enumerate the most relevant non-pharmaceutical interventions, focusing on different research works that assess their efficacy.

\begin{itemize}
    \item \textbf{Mobility restrictions}:  Governments often introduce long-range or local mobility restrictions aimed at reducing disease transmission.   Spatial epidemiology is particularly useful to model the effects of such measures. For instance, in \cite{Brethouwer2020}, the authors propose a social network approach to assess the post-lockdown mobility measurements for Covid-19. A SEIR model is combined with a  small-world network, concluding that the blockage of long-distance mobility can contain the second peak of infected individuals effectively. 
    
\item \textbf{Social distancing}: Social distancing is another measure promoted by governments, public and private institutions in an attempt to reduce disease transmission. Reducing or stopping the activity in educational institutions or factories are examples of this. In \cite{Maharaj2012}, the authors conduct a simulation-based analysis to determine the effects of social distancing both in public health and in the economy. Two social network models (regular and small-world networks) are combined with a compartmental SIR model, and the economic impact takes into account the costs of individuals falling ill and the cost of a reduction in social contacts. The obtained results suggest that social distancing is effective only when adopted in a highly strict manner, giving a worse outcome than doing nothing when implemented in a weak fashion (do it well or not at all). 

\item \textbf{Pro-active testing}: Proactive testing of asymptomatic individuals is very relevant for the monitoring and control of the pandemic \cite{WHO:Cases:and:Clusters}.  
It allows to isolate infected individuals and implement contact tracing strategies which have been shown to be crucial in the effective control of the pandemic \cite{Giordano2020}. 

\item \textbf{Quarantine}: Quarantine of a whole population is the most extreme measure in the scope of social distancing/mobility restrictions. The extreme impact of Covid-19 yield to the quarantine of the epicentre of the pandemic (Wuhan) on January 24th, 2020, and the same measures were subsequently adopted in different countries of Europe and America. In \cite{Dandekar2020}, a SIR model is augmented to consider the time-varying strength of quarantine $Q(t)$. Using data from Wuhan, the authors conclude that SIR and SEIR models (with time-constant parameters) are not able to capture the stagnation of the epidemic caused by the imposed quarantine and need time-varying terms as $Q(t)$ to take into account isolation measurements. Then, $Q(t)$  estimated by means of a deep neural network trained with Wuhan data, which suggests that about 70\% of the infected population was effectively isolated at the peak of the quarantine measures.

\item \textbf{Contact tracing}: Contact tracing is a widely used epidemic control measure that aims to identify and isolate infected individuals as soon as possible, by following the contacts of individuals that are known to be infectious. A review of contact-tracing based epidemic models for SARS and MERS epidemics can be found in \cite{kwok2019epidemic}. In \cite{Kiss2006}, a small-world, free-scale network model is combined with a compartmental model to assess the efficacy of contact tracing.

\item \textbf{Use of masks}: Recommendations and common practices regarding face mask use by the general public have varied greatly and rapidly over the course of Covid-19 pandemic. Messages differ from country to country, and at the same time, some governments or public health bodies were pleading to the population to stop buying masks, other countries were distributing them to the general public \cite{Times2020}. In order to assess the impact of mask use by the general public, a modified compartmental SEIR model is employed in \cite{Eikenberry2020}, taking into account asymptomatic individuals, stratifying the total population into those who habitually do and do not wear face masks in public, and introducing parameters to model mask effectiveness. The model is fitted with data of Washington and New York states around March 2020. Their results suggest that broad adoption of even relative low-quality masks may meaningfully reduce community transmission and decrease peak hospitalizations and deaths.

\end{itemize}

\subsection{Fitting epidemic models to data}\label{subsection:Fitting:to:Data}

Epidemiology dynamical models rely on a set of parameters that have to be tuned in order to provide functional prediction models and/or infer from them essential characteristics, such as the (time-varying) effective reproductive factor \cite{Cori2013}, the latent period, etc. \cite{Giordano2020}. Fitting epidemic models to data is a fundamental problem in epidemiology that can be approached in different ways. A first classification is to distinguish between classical methods, in which the parameters of the model are unknown but fixed, and Bayesian methods, in which they are assumed to be random variables \cite{Kypraios2017}.  Another classification follows from the accessibility to the populations considered in the compartments of the model:
\begin{itemize}
    \item {\bf Full access} to the evolution of the number of cases in each compartment: In most models, the parameters that determine the dynamics enter in a linear way (multiplying linear or bi-linear terms that depend on the current number of cases in each compartment). This means that a (vector) equality constraint, that depends linearly on the parameters to fit, can be obtained at each sample time. Thus, standard linear identification schemes, like least-square methods, can be applied to estimate the parameters that best fit the model to the data. See, for example, \cite[Chapter 6]{Martcheva2015} and \cite{Allman2004}.
    \item {\bf Partial access} to the number of cases in each compartment: In many situations, there are no available time series for one or more of the groups considered in the model. This complicates the data-fitting process considerably because it is no longer possible to obtain, in a simple way, the equality constraints described in the full access case. The standard approach in this case is to resort to non-linear identification techniques (see \cite{Schon2011} and \cite{Abreu2020}). In this context, Monte Carlo based methods (e.g. Markov Chain Monte Carlo and Sequential Monte Carlo algorithms) play a crucial role in addressing the challenges that lie in reconciling predictions and observations \cite{McKinley2009}. 
\end{itemize}



\subsubsection{Sensitivity analysis}

Sensitivity analysis (SA) is the study of how the
uncertainty in the output of a model (numerical or
otherwise) can be apportioned to different sources
of uncertainty in the model input \cite{Saltelli2002:Sensitivity}.
See the review paper \cite{Qian2020:Sensitivity} on the use of this technique in the context of biological sciences. A monovariate and multivariate sensitivity analysis for a data-fitted SARS model is given in \cite{Alvarez2015:MORILLA}. The use of SA is common in many research papers on modelling Covid-19 (see e.g. \cite{Fang2020} and \cite{Saljeeabc3517}). 

\subsubsection{Validation and model selection}

The ultimate test of the validity of any model is that its behaviour is in accord with real data. Because of the simplifications introduced in any mathematical model of a biological system, we must expect
some divergence between the results of a model and reality even for the most carefully collected data and well-constructed
model. Different questions arise: i) How can we determine if a model describes data well? ii) How can we determine the parameter values in a model that are appropriate for describing real data? These questions are much too broad to have a single
answer \cite{Allman2004}, \cite{Vittinghoff2012}.

Epidemic models depend on their data calibration. However, many possible models are potentially suited to analyze the spread of the pandemic in a given moment. The models are inherently linked to the goal for which they were envisaged. For a given goal (for example second outbreak detection), different models can be considered. Model selection techniques are used on a regular basis in epidemiology \cite{Portet2020:AKAIKE:MODEL:SELECTION}. They address the problem of choosing among a set of candidate models the most suitable one \cite{BurnhamKennethPandAnderson2010}. The selection is based on different aspects: i) How the calibrated model is able to reconcile and match observations and ii) The complexity of the model. Under similar adjustment to observations, simpler models are preferred since they are more robust and sound from an information theory point of view \cite{Huyvaert2011:REVIEW:MODEL:SELECTION}. 
  
There are often different sets of parameters yielding a similar fit to data, but providing significantly different estimations of the main characteristics of the spread of the epidemic (like peak size, reproduction number, etc.). This issue is known as nonidentifiability \cite{Roda2020:DIFFICULTIES}. Identifiability issues may lead to inferences that are driven more by prior assumptions than by the data themselves \cite{Lintusaari2016:Identifiability}. There are some approaches to address this difficulty. The first one is to resort to simplified models (SIR and SEIR models, for example) in which the number of parameters to adjust is small \cite{Roda2020:DIFFICULTIES} and \cite{Postnikov2020:BACK:ENVELOPE}. The second one is to use data from different regions in a not aggregated way, which reduces the probability of parametric over-fitting. In this context, model selection theory provides systematic methodologies to determine which model structure best suit the purposes of the model, \cite{BurnhamKennethPandAnderson2010}, \cite{Portet2020:AKAIKE:MODEL:SELECTION}.


\section{Forecasting}\label{sec:forecasting}

Forecasting is a supervised learning approach in which a number of variables or predictors (also called \emph{features} in machine learning literature) are used to predict the value or a category of a variable of interest (target variable). Supervised learning is normally classified into two main categories: i) classification, and ii) regression or forecasting.  Classification methods can be employed for diagnosis and detection of Covid-19 cases (see \cite{Hanumanthu2020} for a review). For instance, to detect Covid-19 cases through X-ray images \cite{khan2020coronet}, \cite{luz2020towards}, \cite{oh2020deep}. In this document, we focus on regression methods and their use to forecast epidemiological variables. The forecasting approaches presented in this section are not necessarily linked to the (prior) design and adjustment of an epidemic model.

The applications of forecasting models related to Covid-19 pandemic are numerous \cite{mahalle2020forecasting} \cite{petropoulos2020forecasting}, ranging from predicting the number of infected cases and deaths, to estimating the parameters of epidemic models such as the rate of infection $\beta$ of a population and the basic reproduction number $R_{0}$. However, some considerations should be taken into account in order to select a suitable model. 
 
 First, from the statistical point of view:
\begin{itemize}
    \item The use of frequentist or classical versus Bayesian empirical statistical methods. In the former, probabilities are assigned according to experiment repetition and occurrence. In the latter, a probability is assigned based on a quantitative understanding of
the nature of the experiment \cite{bonamente2013statistics}, that is Bayes theorem and probability distributions (priors).
    \item Parametric versus non-parametric approaches. In the former, there is a fixed mapping function between the input and the output of the model with several parameters to be obtained. In the latter, there is no such fixed mapping function, or it is unknown. 
\end{itemize}

Second, from the temporal point of view:
\begin{itemize}
\item  Temporal series methods rely on the previous values of the target and the temporal characteristics of the data, such as trend and seasonality, to make predictions. Thus, they are based on autoregression techniques and temporal differences. In contrast, regression models use previous values of the predictors to predict the future value of the target. Furthermore, mixed techniques can be used.
\end{itemize}

Other considerations are:

\begin{itemize}
    \item The model should be trained with reliable data.  If the input data is poor, the forecasts produced will also be poor. Therefore, reliable data should be collected. On this line, techniques such as data reconciliation, standardization, filtering, outlier detection, etc., can improve the raw data collected (see Section \ref{Sec:Consolidated:Time:Series}).
    \item The amount of data for training can vary from one forecast model to another. For instance, deep learning approaches require much more data compared with classical machine learning approaches.
    \item Learning procedures should include training, validation, and test phases executed separately. Therefore, the data set should be divided into three parts, each one used for a different purpose. In the training stage, model parameters are tuned according to the corresponding training data. The validation step is typically used to adjust the model hyper-parameters and to perform comparisons with other counterparts. Finally, the test of the selected model should be carried out with unseen data for reporting the performance of the model. Due to the scarcity of data related to Covid-19, in some cases, the learning procedure can be reduced to training and test.
    
    \item Scalability of selected model with input data. Some models do not scale well with the input data, for instance, kernel-based methods, since the computational burden of their implementation does not grow linearly with the number of observations. 
\item Interpretability of the internal functioning of the models should be considered since it is complicated in some of them. This is the case for instance, in deep learning approaches and complex ensemble methods. It can be questionable to develop decision making for Covid-19 based on models with low interpretability (black box modeling)\cite{arrieta2020explainable}. Other prediction models, like the ones implemented by means of an epidemic model (see Section \ref{sec:models}) are more interpretable because their parameters are directly related to meaningful characteristics of the virus or the considered population. 
\end{itemize}

In the following sections we first list the potential variables that can be used to feed the models. Second, we review the main forecasting tools available to predict future spread of Covid-19 pandemic. We have classified the methods into different categories, such as parametric, non parametric, temporal series methods, and deep learning approaches. Finally, we review different assessing methodologies for validating the models.

\subsection{Identifying relevant variables}

Previous works on epidemic modelling, before the Covid-19 outbreak, already pointed out key features or variables to be considered in successful data-driven forecasting methods, such as demographic variation, mobility patterns that include the entire
global air-traffic system as well as the short-scale, daily commuter movements in almost every country on the planet, detailed epidemiological data, and disease-specific
mechanisms. However,  today,  the availability of big data and IoT technologies makes the list of potential variables to consider much larger. Next, we provide some categories and examples of variables that have been used in the current literature \cite{Alamo2020b}. 
\begin{itemize}
        \item Variables directly associated to the Covid-19 outbreak: Regional time series of the number of confirmed cases, suspicious cases, deaths, recovered, number of tests, hospitalized cases, ICU cases, isolated positive cases, etc.  
        \item Geographic variables: Locations of Covid-19 variables such as GPS coordinates: longitude and latitude.
        \item Demographic variables: Population and density of population. 
        \item Health system variables: Total number of ICU beds, number of doctors and nurses, personal protective equipment (PPE), respirators, number and types of tests.
        \item Government measures: Social distancing, movement restrictions, lockdowns, etc. 
        \item Weather variables: Temperature and relative humidity, among others.
        \item Contamination variables: Air pollution indicators, for example, fine particulate matter $PM_{2.5}$ and $NOx$ concentrations.
        \item Mobility data: International and national mobility patterns. Traffic patterns.
        \item International and national connectivity: Number of international and national flights, number of train connections, etc.
        \item Economic impact variables: Stock market, crude oil, agricultural futures, gold price. \end{itemize}
Furthermore, there are several techniques for identifying relevant variables, also called feature selection in machine learning literature, such as statistical tests (F-test or ANOVA), recursive feature elimination, forward selection, and principal component analysis, among others. In addition, correlation coefficients can be also used to identify potential variables to be considered \cite{gilbert2020preparedness} \cite{Haider2020}.

\subsection{Parametric methods}

These are learning models that condense relations between variables in a set of parameters of fixed cardinality \cite{russel2013artificial}, independent of the number of training examples. The assumptions about the relationship between input and output variables are fixed. For instance, this relation can be linear or exponential. The training of this type of models results on a set of weights (parameters) that determine how each of the inputs affects the model output(s). These weights are obtained through an optimization process that aims at reducing the error between the forecast (model estimations) and the outputs available in the training data. It is important to highlight that, as predicted by epidemiological models (see Section \ref{sec:models}), at the initial stages of the Covid-19 outbreak the number of cases and deaths showed an exponential relationship with time. This fact is used to obtain simple forecasting models at the early stages of the pandemic.

In what follows, we review representative cases for parametric techniques used for fighting Covid-19 pandemic:     
\begin{itemize}
    \item \textbf{Linear Regression:}  
In a linear regression model the estimation of the target output is calculated through a weighted linear combination of the inputs. This is one of the simplest methods for forecasting. The main limitations of linear regression are i) the assumption of linearity between the dependent variable and the independent variables, ii) it is highly sensitive to outliers, and iii) It can be prone to overfitting if there are too many parameters to adjust and not enough data. In the context of Covid-19, linear regression can be used to determine the significance of some variables (weather, demographic, connectivity, etc.) on Covid-19 spread. For instance, the influence of temperature and humidity in the basic reproduction number $R$ in China has been studied in \cite{wang3551767high}.

\item \textbf{Ridge Regression:}
Ridge regression or Tikhomonov regularization is a regularized version of the linear regression aimed at reducing the weights of the model \cite{hoerl1970ridge}. The basic idea is to include a quadratic regularization term to avoid over-fitting. In Ridge regression, an $L_{2}$ norm-based penalty term is added to the cost function to regularize the model. In \cite{Qin2020}, the authors use Ridge regression to predict the number of suspected cases in China using social media information. They use the Baidu Search Index (BSI) to account for social media search indexes (SMSI) such as dry cough, fever, chest distress, coronavirus, and pneumonia. The model uses these variables to predict the number of cases. Moreover, the authors compare the results with other regression models, such as LASSO regression and elastic net. 

\item \textbf{LASSO Regression:}
Least Absolute Shrinkage and Selection Operator Regression (LASSO) is also a regularized version of linear regression \cite{tibshirani1996regression}. In this case, the norm $L_{1}$ is used to penalize the weights of the linear model. The idea is that the weights of the variables that are not important to predict the output often attain a zero value. Regarding its use in Covid-19 pandemic, in \cite{Achoki2020}, the authors use LASSO regression to predict the cumulative cases, new infections, and mortality rates. The model captures a combination of epidemiological, socio-economic and health system readiness related covariates. The results provide predictions for the five sub-regions of Africa (Central, Eastern, Northern, Southern, and Western). The predictions suggest that some variables play an especially important role the spread of the pandemic, such as i) urbanization level and socio-demographic index (SDI), ii) international connectivity iii) living standards and health systems, and iv) population density. 
\item \textbf{Generalized Linear Methods (GLM):}
One of the main assumptions of linear models is that the error in the prediction follows a normal distribution. However, this assumption does not hold in many scenarios, for instance, in the case of binomial or Poisson responses \cite{myers1997tutorial}\cite{Zhang2020}. The GML methods use link functions to generalize the response of regression models. As application cases on Covid-19, in \cite{Kraemer2020}, the authors use GLM to predict the growth rate of Covid-19 in China as a function of mobility. A similar approach can be found in  \cite{wu2020generalized}, where the authors use Generalized Richards model (GRM) \cite{richards1959flexible} to predict the number of cases in 29 provinces of China. Another example of GLM can be found in \cite{Roosa2020}. In \cite{Saez2020}, the authors use a generalized linear mixed model (GLMM) with a variable response from the Gaussian family to predict the number of accumulated cases in Spain. Moreover, in \cite{shi2020impact}, a Generalized Additive Model (GAM) is employed to estimate the effect of temperature and the spread of Covid-19 in China. In \cite{WU2020139051}, the study is extended for 166 countries. A Poisson model is used in \cite{Zhang2020} to detect turning points in the epidemics of some western countries such as Canada, France, Germany, Italy, UK and USA. 

\item \textbf{Gaussian error function:}
The Gaussian error function or \textit{erf} is a sigmoid function used in probability, statistics and partial differential equations to describe diffusion phenomena. This function can be used for predicting the exponential behaviour of some Covid-19 variables, such as the number of new cases and/or new deaths. However, one important limitation of this approach is the fact that it is based on curve fitting of some parameters and, therefore, the quality of the forecasting depends on both the data and the fitting method. This is particularly critical at the early stages of the outbreak because of time-varying criteria for confirmed-cases. In \cite{Team2020}, the authors use a Gaussian error function to predict the rate of deaths. They used curve fitting and data from Wuhan. The model is used along with a microsimulation model to predict health system utilization in USA in the following four months. A similar approach for forecasting the number of positive cases in Italy can be found in \cite{Ciufolini2020}. 
   
\end{itemize}

In the context of parametric methods, it is important to make reference to Bayesian inference, which makes it possible to incorporate prior information at the early stages of an epidemic \cite{Dehningeabb9789}, updating the predicted outputs with posterior probabilities as more data is provided. In addition, Bayesian methods are more suitable for quantifying uncertainties than classical frequentist methods \cite{Dehningeabb9789}. There exist a large number of Bayesian tools \cite{gelman2013bayesian}, such as linear regression, hierarchical models, generalized models, Gaussian processes, and Dirichlet process, among others. As an application case in Covid-19 pandemic, in \cite{Flaxman2020} the authors use a semi-mechanistic Bayesian hierarchical model to estimate the impact of government measures on Covid-19 spread in 11 European countries. By using the model, they assess the mitigation actions taken by 11 European countries. Because of the scarcity of data, the model is significantly affected by countries with more cases, such as Italy and Spain. Furthermore, Bayesian inference has been widely used to incorporate prior knowledge and uncertainty in the parameters of epidemic models \cite{Ferretti2020} \cite{Saljeeabc3517} \cite{Dehningeabb9789}. As limitations of Bayesian methods, the selection of priors can be difficult in some scenarios since it depends on previous experience on the field. 

\subsection{Non-parametric approaches}
In non-parametric techniques, there is not any predefined mapping function between the input variables and the target output. In other words, no assumption is made about the relation between the features and the forecast output(s). In this section, we review the most significant non-parametric methods that can be used for forecasting in Covid-19 related problems. Additionally, some related works are briefly described.

\begin{itemize}
    
\item \textbf{Kernel methods}:
These techniques are based on the ``kernel trick", that is, using mapping functions to transform inputs in a new dimensional space, normally of a much higher dimension. Nonlinear forecasting problems can be transformed into high dimensional ones by using appropriated kernels. Similarity functions are usually suitable kernels like the well-known Gaussian Radial Basis Function (RBF). Support Vector Machines (SVM) are by far the most representative techniques of kernel methods \cite{cherkassky2004practical}. SVM are widely used for both classification and regression tasks. The main limitation of kernel-based methods is that they do not scale well with large data sets since the quadratic problems involved in their numerical implementation grow with the number of data points \cite{pavlov2000towards}. 
 
\item \textbf{Symbolic regression}:
In symbolic regression, the objective is to achieve the optimal relationships/operations, in the form of a relational tree, between the input variables in order to reduce the prediction errors \cite{davidson2003symbolic}. Since no assumption is made between the predictors and the target, a set of possible operators, constants, and analytic functions are predefined. Thus, the optimal operations among the variables and their order are normally obtained through an evolutionary computation approach \cite{koza1994genetic}. With respect to machine learning approaches, in symbolic regression, both model structures and model parameters are optimized. As a result, a tree of operations among the predictors is obtained. As an example of application, in \cite{Salgotra2020} a genetic programming approach is used to predict the the number of cases and deaths due to Covid-19 in India. The main limitations are: i) prone to overfitting and ii) poor interpretability. Both issues can be alleviated by limiting the depth of the resulting trees. 
\end{itemize}

\subsection{Deep Learning}

Deep Learning techniques represent an evolution of Artificial Neural Networks (ANNs), where the number of layers is increased notably \cite{Goodfellow-et-al-2016}.  A standard deep neural network (DNN) is, technically speaking, parametric if it has a fixed number of parameters. However, most DNNs have so many parameters that they could be interpreted as non-parametric. 
Deep learning neural networks are sequentially improved using successive slots of training data. Thus, these networks require much more data than classical machine learning techniques. One of the main features of deep learning consists in the determination of an encoding vector from the raw input, features or variables. Therefore, they do not need feature engineering of the predictors. Each layer contributes to the encoding vector: the first ones obtain basic features of the data, while deepest layers use the previous results to develop more complex and higher-level features. Several well-known architectures of deep learning networks can be found depending on the structure and type of data that is used as input, such as Fully-Connected Networks (FCN), Recurrent Neural Networks (RNN) and Convolutional Neural Network (CNN).  
\begin{itemize}
    \item \textbf{Fully Connected Networks}: In FCNs, each neuron of a given layer is connected to all the neurons of the previous layer. In \cite{Rizk-Allah2020}, the authors use a feed-forward neural network improved with an interior search algorithm to predict the number of cases of Covid-19. Similarly, in \cite{Torrealba-Rodriguez2020} the authors use a three-layer network to predict the number of cases in Mexico. In \cite{baltas2020monte}, a FCNs is used with a SIR model to predict the peak of Covid-19 in Spain. 
    
    \item \textbf{Recurrent Neural Networks}: RNNs are the main tool for studying data in the form of temporal series in the deep learning community. In RNNs, the neurons of a given layer are fed with both the previous layer outputs and previous values of the following layers. Therefore, bidirectional communication is established at each layer. This characteristic allows determining short and long terms relationships among the target output and the predictors. One advantage of RNNs with respect temporal series methods is that RNNs do not require to remove trend and seasonality (although it can improve the obtained results). Regarding the use of RNN for Covid-19 analysis, in \cite{Yang2020}, the authors use an RNN to predict the number of new infections. They trained the network with data from the 2003 SARS epidemic and then used the results in a SEIR model to forecast epidemics trend of Covid-19 in China under public health interventions. In \cite{chimmula2020time}, a Long Short Term Memory (LSTM) is used to forecast the number of cases in Canada outbreak. A similar work can be found in \cite{TOMAR2020138762}, where an LSTM architecture is used to estimate the number of cases in India. In \cite{zheng2020predicting} a LSTM model and NLP techniques are combined to predict Covid-19 propagation in China.

    \item \textbf{Convolutional Neural Networks}: CNNs are in general suitable for image classification tasks, in particular 2D CNNs. However, they can also be used for temporal series like 1D CNNs. CNNs apply convolutional filters to the input data to obtain an encoding vector of the data. In the context of Covid-19, 2D CNNs can be used for detection and diagnosis \cite{khan2020coronet}, \cite{luz2020towards}, \cite{oh2020deep}. The main idea behind these works is to present alternative techniques for RT-PCR testing. Another interesting research direction to explore is 1D CNNs. These networks have been successfully employed in data processing tasks \cite{li2017classification}. In 1D CNNs, convolutional filters are applied to several temporal series simultaneously. With respect to RNNs, the 1D CNNs are suitable for obtaining short term relationships among the input and the output, while RNNs allows long terms explanations.
\end{itemize}

 Finally, in \cite{Apostolopoulos2020}, the authors highlighted the potential of transfer learning in deep learning approaches. This technique is especially useful when data is relatively scarce, as it is the case in Covid-19 pandemic. The idea is to transfer pre-trained architectures successfully employed in forecasting problems to make predictions of variables related to Covid-19. 

The main limitations of deep learning approaches are: i) they require a high amount of data, ii) hiperparametrization of the models is complex, and  iii) the interpretability of the resulting model is low.

\subsection{Ensemble methods}
These methods are based on the aggregated response of several forecasting models. Also known as \textit{wisdom of the crowd} \cite{sagi2018ensemble}, the main idea behind these methods is that a suitably combined prediction from many models will hopefully outperform the predictions of the best one. The diversity of the models in the ensemble plays an important role since each of them can specialize in learning some input-output relations or aspects of the predicted output. As a representative example of ensemble methods, Random Forest technique \cite{liaw2002classification} aggregates the response of many decision trees. The main limitations of these techniques are: i) an increase of the complexity of the model with respect to single models and, thus, an increase in the amount of data required for training, and ii) reduced model  interpretability. An ensemble method is proposed in \cite{benitez2020short} to obtain one-week ahead forecasts of the number of Covid-19 hospitalized cases in Andalusia, Spain. The predictors used are SVM regressors and Random Forests, and a blending method is used to aggregate the responses. In \cite{cobb2020examining}, the authors use random forest to evaluate social distancing on the compound growth rate of Covid-19 at the county level in the USA. In \cite{ribeiro2020short}, an ensemble method is used to predict the number of cases in Brazil, the techniques used are the cubist, random forest, RIDGE, and SVR as base-learners, and Gaussian process (GP) as meta-learner.

\subsection{Time Series Theory}

 Time series analysis comprises techniques for analyzing time series data. Time series models used for forecasting include decomposition models, exponential smoothing models and ARIMA models, among others (see \cite{hyndman2018forecasting}). One of the main differences with regression models is that the predictors are previous values of the target variable. For instance, if we consider the number of new confirmed cases of Covid-19 cases as a temporal series, the forecasting method uses the previous values of new confirmed cases to predict future values. Normally, linear models are considered; however, these methods can be extended to non-linear relationships.  
 In what follows, we review different techniques from the field of time series theory, that are used to forecast Covid-19 spread: 
\begin{itemize}
    \item \textbf{Decomposition models}: A time series can be divided into three components: a trend-cycle or trend component (T), a seasonal component (S), and a remainder component (R). Thus, the target value to be predicted can be expressed as $y_{t}=S_{t}+T_{t}+R_{t}$ (multiplicative formulations can also be used). Examples of decomposition techniques are X1, SEATS, and STL (see Section 6 in \cite{hyndman2018forecasting}). These methods can be used to determine the trend and seasonality of Covid-19 cases.  
    \item \textbf{Auto Regressive Moving Average (ARMA and ARIMA)}: This approach combines Auto Regressive (AR) and Moving Average (MA) models. AR model specifies that the target variable depends linearly on its own previous values plus a random white noise error. In MA model, the output is the sum of the mean of the target variable and a linear relationship of the previous white errors. MA is used to measure the trend component. The ARIMA model is a generalization of ARMA model, which is suitable for non-stationary data. Thus, both trend and seasonality of the time series change over time. The integrated part of the model is employed to eliminate the non-stationary. Notice that Covid-19 spread changes over time due to contention and mitigation techniques employed by decision-makers. Therefore, ARIMA models are suitable for Covid-19 forecasting. As application examples, in \cite{benvenuto2020application}, the ARIMA model is used to forecast the spread of the virus in terms of prevalence and incidence. In \cite{chintalapudi2020covid}, an ARIMA model is used to estimate the number of cases and recovered in Italy after several days in a lockdown scenario. Similarly, in \cite{ceylan2020estimation} prevalence is estimated in France, Italy and Spain through an ARIMA model. Furthermore, there are improved ARIMA models, for instance, in \cite{Ahmar2020} use a SutteARIMA model to predict stock market in Spain. 
\end{itemize}

Regarding the limitations of time series methods, they work well for measuring trends and seasonality of temporal data; however, since these methods rely on past data (past events), they present difficulties for modelling the effects of events that have not happened before. For instance, in Covid-19 pandemic context, they can be used for determining the trend and seasonality before and after control techniques are employed. Nevertheless, during the transition phases, these methods could present some problems to estimate the changes in a short period of time. Furthermore, the poor quality of the data reported on Covid-19 temporal series, for instance, changing criteria (see \cite{Alamo2020b}) can also be an issue for time series methods.    

As for a comparison work between temporal series techniques and other methods, see for example \cite{kane2014comparison} where a comparison of ARIMA model and random forests for an influenza pandemic in Egypt is presented.


\subsection{Assessing the performance of forecasting models}

The generalized performance of a forecasting method can be defined as its prediction capacity on an independent test, that is, data that the model has not seen previously. The objective of assessing methodologies is to reduce error and avoid overfitting, i.e., the model works well during training but poorly during testing. Therefore, assessment methods are crucial to validate the developed forecasting methods. The validation process in the context of the Covid-19 pandemic is not simple because available data can be highly correlated. In order to address this issue, it is desirable to validate the forecasting methodologies with data sets from different regions and different periods of time. This is much easier to be done with other infectious diseases like seasonal influenza because of the availability of time-series from different locations and different years.

The literature reviewed to prepare this paper shows that the vast majority of proposed models for Covid-19 forecasting do not include suitable assessment techniques. The models are normally trained with data of similar specific locations. Such models perform relatively well in these locations. However, they might not generalize well for other territories or periods of time. Besides, often just one single model or methodology is proposed, without comparing the performance with other counterparts.

The assessment of models includes the validation and test phases. The main objectives are:
\begin{itemize}
    \item Determining the optimal hyper-parameters for a given prediction model. This step is also called hyperparametrization and it is crucial for a fair comparison among different models. This procedure should be employed in the validation step \cite{snoek2012practical}. 
    \item Model selection by the comparison of several models.  Model selection should also be carried out in the validation phase \cite{Portet2020:AKAIKE:MODEL:SELECTION}.
    \item Assessment of the performance of the model by estimating the generalization error of the selected model using new data in the test phase.
\end{itemize}

Assessment could be performed including data of a collection of territories with similar characteristics (See Cluster methods in Section \ref{Sec:Consolidated:Time:Series}) in terms of the state of the pandemic, size, population, weather, etc. 
The two main techniques used for evaluating the generalized error are cross-validation and bootstrapping.
\begin{itemize}
\item \textbf{Cross validation}:
It is a useful tool used for model validation in cases of a scarcity of data \cite{friedman2001elements}.
In cross-validation \cite{BERGMEIR2012192}, the whole data is used for both training and validation. In general, cross-validation resamples without replacement and thus produces surrogate data sets that are smaller than the original. As an example, K-fold cross-validation consists of dividing the data randomly in k subsets of approximately equal sizes. The model is trained with \textit{k-1} parts and validate the resulting model with the remaining part. This procedure is repeated $k$ times, varying each time the parts used for training and validation. Then, the cross-validation estimate of
prediction error is calculated by averaging out the results. Suitable values for k are ranging from 5-10 \cite{friedman2001elements}. Cross-validation is widely used for hiperparametrization of model and comparison purposes. As an example of usage of cross-validation see \cite{Achoki2020}.
\item \textbf{Bootstrapping:}
The basic idea of bootstrapping is
to randomly draw data sets with replacement from training data, where each sample has the same size as the original training set \cite{friedman2001elements}. To validate the generalization capacity of a model, a suitable bootstrap is the leave-one-out approach, where predictions are made from bootstrap samples not containing that observation from the training data. Bootstrapping is used to characterize, in a statistical way, the errors produced by the forecasting method. 
\end{itemize}
\subsubsection{Performance metrics}

Several criteria can be considered to measure the performance of forecasting models. Let us consider $n$ data samples, and denote with $y_{i}$ the actual outcome of the variable to be predicted ($i=1,...,n$), and with $\hat{y}_{i}$ the value predicted by the forecasting model. The following performance metrics can be defined:

\begin{itemize}
    \item \textbf{Mean Absolute Error (MAE)}: It is calculated as the arithmetic average of the absolute error. MAE is an absolute measure of fit.
\end{itemize}
\begin{center}
$MAE=\fracg{1}{n}\Sum{i=1}{n}|y_{i}-\hat{y}_{i}|.$ 
\end{center}

\begin{itemize}
    \item \textbf{Mean Square Error (MSE)}: It is calculated as the average squared difference between the estimated values and the actual value. MSE is more sensitive to outliers than MAE since the error is squared. A common variant is the root value of MSE or RMSE, which can be interpreted as the standard deviation of the unexplained variance.
\end{itemize}

\begin{center}
     $MSE=\fracg{1}{n}\Sum{i=1}{n}(y_{i}-\hat{y}_{i})^{2}.$
\end{center}

\begin{itemize}
    \item \textbf{R-squared}: It measures the proportion of the variance for a dependent variable that is explained by an independent variable or variables. $R^{2}$ values are within the interval $[0, 1]$ and a value close to one is desired. A common variant is the adjusted R-square, which is suitable for models with high number of independent variables.
\end{itemize}
\begin{center}
     $R^{2}=1-\fracg{\Sum{i=1}{n}(y_{i}-\hat{y}_{i})^{2}}{\Sum{i=1}{n}(y_{i}-\mu_y)^{2}},$
\end{center}
where $\mu_y$ denotes the empirical mean value of $y_i$, $i=1,\ldots,n$.
\begin{itemize}
    \item \textbf{F-test}: It compares the proposed model with a model with no predictors. Thus, it is used to reject the null hypothesis. A high value is desired.
\end{itemize}

Other metrics can be used. However, the previous ones cover the main spectrum. In addition, each performance metric evaluates a model in a different way, and the suitability of each depends on the data and the selected model.

\section{Impact Assessment Tools}\label{sec:Impact:Tools}
In order to be able to design effective control strategies, it is important to define the goals first. 
It is relatively easy to formulate the objectives in a qualitative way: the ultimate goal is to maintain the spread of the virus within an adequate threshold while minimizing the economic and social impacts of the interventions. Possible non-pharmaceutical control strategies include social distancing, border closures, school closures, isolation of symptomatic individuals, among others.

However, to resort to advanced decision-making techniques, like the ones proposed in the next section, one has to state the control objectives in a more precise way. Furthermore, it is necessary to assess which interventions have the desired impact of controlling the epidemic and, besides, which of those or other actions are necessary to maintain control \cite{Cauchemez2006:Assessment:efficacy}. 

This tasks can be accomplished by employing suitable indexes that model the impact of the actions on the different aspects considered in the decision-making process. Examples of possible indexes are the mean reproductive number, the mortality index, the unemployment rate or public debt, to name just a few. 
Once the indexes are established, the goals can be stated in a formal way in terms of constraints and minimization of a weighted cost index. For example, the goal could be stated in terms of satisfying that the mean reproductive number is kept smaller than one while minimizing a weighted index modelling the impacts on economy and society. 

The quantitative computation of such indexes is not an easy task because they require the design of data-driven strategies to assess the effect of the decisions on the different indexes. This can be done by means of the predictive models and forecasting schemes analyzed in the previous sections. 

In some cases, to demonstrate a significant effect of one measure or intervention over the spread of the epidemic is complex since multiple factors are present simultaneously. Thus, the isolation of one parameter over others is difficult or impossible \cite{OHANNESHAUSHOFER2020}. In these scenarios, correlation analyses like Pearson Correlation Coefficient (PCC) can be a naive way to assess the effect \cite{Haider2020}. A more interesting approach to conduct these analyses is the so-called Randomized Control Trials (RTC) \cite{OHANNESHAUSHOFER2020}. In an RTC, a subset of randomly chosen individuals or regions receives an intervention and randomly selected control groups receives no intervention. This enables to evaluate the impact of the intervention taken.

We now enumerate, for the most relevant aspects of the Covid-19 pandemic context, possible indexes that could be included in the decision-making process. 

\subsection{Spread of the virus and reproductive number}
It is natural to express the effectiveness of control strategies in terms of the effective reproductive number $R(t)$. As introduced in Section \ref{sec:models}, the basic reproduction number $R_{0}$ determines the potential of an epidemic to spread exponentially at its early stage. In contrast, when an epidemic is on course, the effective reproduction number, $R(t)$, is a time varying quantity that indicates the average number of secondary cases produced by an infectious individual. The effective reproduction number can assess the ability of control measures to decrease the spread of an epidemic. By combining measures that maintain $R(t)$ below 1, the incidence of new infections decreases and the spread of epidemics fades with time. 

In \cite{Cori2013}, the authors presented a software tool that can be executed with Microsoft Excel or R. The tool was validated with 5 different epidemics, including SARS and influenza, being able to estimate the daily reproductive number $R(t)$ and its variation in the presence of vaccination and super spreading events. 

For Covid-19 pandemic, an example of this kind of assessment can be found in \cite{Fang2020}, where using data and a SEIR model, the authors estimate $R(t)$ in Wuhan city and conclude about the effectiveness of government measures. However, the dynamic trend of $R(t)$ depends on a parameter $k$ which role and definition is unclear. Based on the number of deaths, in \cite{Flaxman2020}, the Imperial College Covid-19 Response Team used a semi-mechanistic Bayesian model to estimate the evolution of $R(t)$ when measures as social distancing, self-isolation, school closure, public events banned, and complete lock-down are recommended/enforced. 

Limitations of using $R(t)$ as an assessment tool arisen from the problems presented in Section 2 related to the data sources. The procedures used to collect the data about infections and deaths are far from being reliable. As a result, determining the real value of $R(t)$ is difficult. Other indirect measures, like the number of deaths, ICU cases, saturation of health-care systems are also employed to assess the current epidemic burden. See the next subsection.

\subsection{Saturation of the health-care systems}

From the early stages of Covid-19 epidemic, its virulence and high contagiousness collapsed or made insufficient, health-care resources in different places around the world, resulting in higher mortality rates \cite{Ji2020,Lai2020}. Furthermore, in countries with low capacity like African and South American countries saturation levels are reached even with a significantly smaller number of cases \cite{Achoki2020}. 

The preparedness of each country to respond against a disaster like Covid-19 pandemic is a useful metric to measure the capacity of each country. This is especially relevant to account for the initial situation of each country to fight Covid-19. The IHR MEF is used as a metric to account for preparedness in \cite{gilbert2020preparedness}. This is a  set of four components  developed  by  WHO  to  support  the  evaluation  of  a  country’s  functional ability to detect and respond to a health emergency. The  IHR  MEF  is composed of a mandatory self-reporting (SPAR report)   of capacity and three optional components, such as the Joint External Evaluation, the after-action reviews,  and simulation exercises. The SPAR  report develops five indices:  (1)  prevent, (2)  detect,  (3)  respond,  (4)  enabling function,  and  (5) operational readiness. On this line, in \cite{kandel2020health} IHR MEF metric is used to evaluate in the context of Covid-19 the health system capacity of 182 countries. 

To limit the saturation of health-care systems and plan the distribution of these resources, it is of invaluable help to count on tools to assess how the different interventions have an effect on the magnitude and timing of the epidemic peak during first and secondary outbreaks (see Sections \ref{sec:models} and \ref{sec:forecasting}). However, obtaining precise tools to forecast these peaks is challenging due to the limitations of the available data and the time-varying nature of the mitigation efforts and potential seasonal behaviour of the pandemic. In order to circumvent these issues, forecasts of cumulative disease burden are often looked for. While missing the intensity and timing of the peaks, these projections can at least allow to identify areas with heavy present and/or future affection of Covid-19 \cite{Miller2020}. 
\subsection{Social impact}
With regard to assessment tools, it is really difficult to aggregate all social aspects affected by Covid-19 pandemic in just one metric. Here we review some of the methodologies that could be helpful to design indexes to model the social impact of the pandemic. 

\begin{itemize}
    \item \textbf{Social network analysis}: social networks such as Facebook and Twitter can be used to assess the impact of Covid-19 in society. People post in these social networks, their feelings and worries. In \cite{shanthakumar2020understanding}, 530.206 tweets in the USA were analyzed to measure the social impact of Covid-19. The hashtags were classified into six categories like general covid, quarantine, school closures, panic buying, lockdowns, frustration and hope. Thus, the number of tweets in each category can be used as a metric of impact. Similarly, Weibo microblogging  social network was used in \cite{li2020characterizing} to study the propagation of situational information on social media related to Covid-19 in China.
    \item \textbf{Search engines}: the searches made by citizens in search engines, such as Google, Bing, and Baidu, among others, can be used to measure the social impact of the epidemic in different locations. Normally, people try to find information of unknown diseases, drugs, vaccines, and treatments on the Internet. On this line, it has been demonstrated correlation of relative frequency of certain queries in Google and the percentage of physician visits in which a patient presents with influenza-like symptoms \cite{Ginsberg2009}. Furthermore, other works have studied this assessment tool for other epidemics like Influenza Virus A (H1N1) pandemic \cite{Cook2011}. Regarding Covid-19, in \cite{Qin2020}, the Baidu engine is used to estimate the number of new cases of Covid-19 in China by the number of searches of five keywords, such as dry cough, fever, chest distress, coronavirus, and pneumonia. These five keywords showed a high correlation with the number of new cases. 
    \item \textbf{News}: the number and the content of posts in online newspapers can also be used to assess the spread of the virus. On this line, in \cite{zheng2020predicting}, Natural Language Processing (NLP) technique is used to extract the relevant features of news media in China.
    \item \textbf{Online questionnaires}: Another tool for measuring the social impact of Covid-19 is through online questionnaires such as \cite{oliver2020covid19impact} (Spain,  146.728 participants) \cite{qiu2020nationwide} (China, 52.730 participants) and \cite{kleinberg2020measuring} (UK, 2500 participants). These allow to rapidly formulate to citizens multiple questions related to psychological, social and economic impact, among other aspects. The main difficulty is to spread the questionnaires throughout the population, although social networks and web-based tools help to reach a large amount of population.
    \item \textbf{Mobility}: reduction of mobility is not only due to the imposed quarantines and lockdowns by governments. Such reductions is also explained by the increasing population's fear of getting infected. In \cite{Engle2020}, the perceived risk index of contracting Covid-19 is defined. This metric is measured the individuals’ perception of risk, and it is determined by several variables, such as prevalence in both local and neighbouring locations, as well as population demographics. The results in \cite{Engle2020} indicate that a rise of local infection rate from 0\% to 0.003\% reduces mobility by 2.31\%. 
\end{itemize}

\section{Decision Making}\label{sec:Decision:Making}


Determining which of the far-reaching social and economic restrictions are most effective and the conditions under which they can be safely lifted is one of the main goals of data-driven decision approaches to combat Covid-19.   Unlike an unmitigated pandemic, which burns through the susceptible population and eventually fades out, a mitigated first wave preserves a population of unexposed, susceptible individuals. This means that when social distancing guidelines are relaxed, the epidemic can once again spread worldwide. It is compulsory to put in place surveillance systems and reactive mechanisms to reduce the potential burden of secondary epidemic waves. The decision-making process in this context is complex for many reasons:
\begin{itemize}
    \item The uncertainty on some crucial parameters characterizing the spread: seasonality, extent and duration of immunity, etc. \cite{Cobey2020:SEASONAL}, \cite{Kissler2020:POST:PANDEMIC}.   
    \item The difficulties in assessing the quantitative effect of a specific set of mitigation interventions on the effective reproduction number \cite{OHANNESHAUSHOFER2020}. 
    \item The significant non-symptomatic transmission of Covid-19, which render some interventions less effective \cite{Nishiura2020:Serial:Interval}, \cite{Prathereabc6197} .
    \item Its different regional incidence, which motivates decisions in a spatially distributed way \cite{Selley2015}.
    \item The necessity of an adequate trade-off between mitigating the spread of the epidemic and reducing the socio-economic impacts. 
    \item The time-delay nature of the system, which does not allow for a prompt evaluation of the effect of the implemented actions. 
    \item The difficulties of assessing in a quantitative way the disruptive effects of the undertaken measures in relevant macroeconomic variables.
    
\end{itemize}

In what follows, we analyze under which circumstances the epidemic can be mitigated (controllability of the pandemic). After that, we also discuss some methodologies that have been applied to combat other infectious diseases, or that could potentially be applied in the context of Covid-19 pandemic. See also the review paper \cite{Nowzari2016:ieee:CONTROL} for the use of control theory in the context of pandemic control. 

\subsection{Controllability of the pandemic}

In this subsection, we review the most important factors determining the controllability of the pandemic. That is, those aspects that have a relevant impact on the effective reproduction number. We link them with standard epidemic threshold theorems (eg. \cite{Becker1977}, \cite{Whittle1955}, \cite{kermack1927contribution}). 

The epidemic threshold theorem of Kermack and McKendrick \cite{kermack1927contribution}, stated in 1927, and in particular, its stochastic form as given by Whittle \cite{Whittle1955}, is fundamental to predict the size and nature of an outbreak of an infectious disease. The
theorem indicates that in homogeneously
mixed communities major epidemics can be prevented by keeping the product of the size of the susceptible
population, the infection rate, and the mean duration of the infectious
period, sufficiently small \cite{Becker1977}. 
We now discuss how to have an impact on each of these factors by means of control actions.

\begin{itemize}
\item {\bf Size of the susceptible population: } The most effective way to reduce the susceptible population is by means of vaccines, which are able to increase the herd immunity to a level that prevents further spread of the disease. A key question is whether protection against Covid-19 will happen by widespread deployment of an effective vaccine or by repeated waves of infection over the next few years until 60\%
to 70\% of people develop immunity  \cite{Graham2020:VACCINE}. Another issue is the duration of the acquired immunity \cite{Kissler2020:POST:PANDEMIC}, which in some infectious diseases, like the seasonal influenza, is not long enough to prevent recurring seasonal peaks \cite{Cobey2020:SEASONAL}.   
\item {\bf Infection rate: } This factor can be reduced by means of different control actions like social distancing, mobility constraints, prohibition of certain activities, etc. \cite{Park2020:REproductive:Number}, \cite{Kraemer2020}, \cite{Ngonghala2020:control:assessment}. It is suspected that Covid-19 will exhibit a seasonal behaviour,  like influenza viruses and the four seasonal human coronaviruses HKU1, NL63, OC43 and 229E \cite{Cobey2020:SEASONAL}, \cite{Neher2020:Seasonal}. Depending on the seasonality and the specific demographic characteristics of a given population, the implemented measures can exhibit a time-varying effect on the infection rate \cite{Cori2013},\cite{Fang2020}. This might cause flows from tropical to temperate regions and back in each hemisphere’s respective winter,  limiting opportunities for global population declines \cite{Cobey2020:SEASONAL} and implying that surveillance methods to detect a seasonal peak should be put in place. 
 
\item {\bf Mean  duration of the infectious period: } 
An effective way to reduce the infectious period consists in detecting infected cases and setting them into quarantine. 
However, because of the relative small latent period and the asymptomatic cases, the impact of this measure depends very much on how fast the detection is taking place. It has been shown that the probability of control decreases with long delays from symptom onset to isolation \cite{Hellewell2020}, \cite{Ferretti2020}. This is one of the key issues because of asymptomatic cases and the significant probability that transmission occurs before the onset of symptoms (median latent delay is slightly smaller than median incubation time) \cite{Ferretti2020}. 
\end{itemize}



\subsection{Optimal allocation of limited resources}
Among other resources, Covid-19 pandemic has provoked shortages in intensive care beds, ventilators, tests, high-filtration masks and other Individual Protection Equipment (IPE). This fact led to the problem of how to ethically and consistently allocate resources. In this context, the term ``resource'' is not limited to hospital beds or ventilators, but extends itself to issues as where and when to allocate the available resources.

A rigorous and precise allocation method should lead to the formulation of an optimization problem, composed of a mathematical formulation and efficient algorithms to obtain its numerical solution. In the mathematical model, resource allocations are the decision variables while the objectives are encoded in cost functions and equality or inequality constraints. For example, in \cite{Zaric2002} and \cite{Brandeau2003}, budget allocation models for multiple population are provided. In \cite{preciado2013optimal},  a network model is used together with geometric programming in order to optimally allocate resources to eradicate an initial epidemic outbreak (see also \cite{Enyioha2015:Resoruce:Allocation}). The same authors extend this last result to the optimal allocation of vaccines in \cite{Nowzari2017}.

However, to the best of the author's knowledge, the literature addressing the optimal allocation of resources specifically for Covid-19 is scarce. Some exceptions to this include \cite{Lampariello2020}, where an optimization problem is formulated to find the number of tests to be performed in the different Italian regions in order to maximize the overall detection capabilities. The problem is a quadratic, convex optimization programming. In \cite{Gollier2020}, a group testing approach is considered, and it is shown how the optimization of the group size can save between 85\% and 95\% of tests with respect to individual testing.

Estimation, forecasting, and impact assessment techniques, discussed in previous sections of this work, are often used to decide the allocation of resources, as they enable decision-makers to predict imbalances between supply and demand and to evaluate the overall efficiency of different alternatives of allocation.

In \cite{Emanuel2020}, the authors propose fair resource allocation guidelines in the time of Covid-19. These guidelines come from 4 fundamental values: i) maximizing the benefits, ii) treating people equally, iii) promoting instrumental value, and iv) giving priority to the worst off. As a result, these guidelines are condensed in some recommendations:

\begin{enumerate}
    \item To maximize the number of saved lives and live-years, with the latter metric subordinated to the former.
    \item To prioritize critical interventions for those health care workers and others who take care of ill patients because of their instrumental value.
    \item For patients with similar prognoses, equality should be invoked and operationalized through random allocation.
    \item To distinguish priorities depending on the interventions and the scientific evidence (e.g. vaccines could be prioritized for older persons while allocation ICU resources, depending on prognosis, might mean giving priority to younger patients).
    \item People who participate in research to prove the safety and effectiveness of vaccines and therapeutics should receive some priority for interventions.
\end{enumerate}

\subsection{Trigger Control}

We comment on this section a strategy that permits to modulate the intensity of the non-pharmaceutical interventions. 
The idea is to implement a trigger mechanism to maintain the effective reproductive factor close to one, avoiding the saturation of the health care system while reducing, when possible, the economic and social burden of the pandemic \cite{Cauchemez2006:Assessment:efficacy}.  
The on-line surveillance of the pandemic permits to estimate the time-varying value of the effective reproductive factor. Three cases are possible:
\begin{itemize}
    \item The effective reproductive factor is clearly under 1: in this case, one could consider lifting one, or more non-pharmaceutical measures. However, other criteria should be met in order to implement a reduction on the confinements measures in a safe way. The three criteria highlighted by the European Commission to decide on the lifting of confinement measures \cite{EuropeanCommission2020:Lifting} are: 
\begin{enumerate}
    \item Epidemiological criteria showing that the spread of the disease has significantly decreased and stabilised for a sustained period of time. This can, for example, be indicated by a sustained reduction in the number of new infections, hospitalisations and patients in intensive care.
    \item Sufficient health system capacity, in terms of, for instance, occupancy rate for Intensive Care Units; adequate number of hospital beds; access to pharmaceutical products required in intensive care units; reconstitution of stocks of equipment; access to care, in particular for vulnerable groups;  availability of primary care structures, as well as sufficient staff with appropriate skills to care for patients discharged from hospitals or maintained at home and to engage in measures to lift confinement (testing for example). This criterion is essential as it indicates that the different national health care systems can cope with future increases in cases after lifting of the measures. At the same time, hospitals are likely to face a backlog of elective interventions that had been temporarily postponed during the pandemic’s peak. Therefore, states’ health systems should have recovered sufficient capacity in general, and not only related to the management of Covid-19.
    \item Appropriate monitoring capacity, including large-scale testing capacity to detect and monitor the spread of the virus combined with contact tracing and possibilities to isolate people in case of reappearance and further spread of infections. Antibody detection capacities, when confirmed specifically for Covid-19, will provide complementary data on the share of the population that has successfully overcome the disease and eventually measure the acquired immunity.
\end{enumerate}
    \item The effective reproductive factor has increased to a level clearly above 1: this would demand, in most cases,  prompt strengthening of the mitigation interventions. The intensity of the new measures should guarantee that the health system is not overwhelmed with the arriving new epidemic wave. This requires the implementation of forecasting tools that help to decision-makers to determine the most suitable set of mitigating measures. 
    \item The effective reproductive factor is close to 1: in this case, a deeper analysis is required. The decision on whether to keep the same set of current mitigation measures or not will depend on the current fraction of infected population, the health-care system capacity, and the potentiality of implementing in a short period of time a mitigating intervention, which is capable of bringing the effective reproductive number to admissible values. That is, the decision could be determined by the worst-case cost of delaying in one week the implementation of new measures.   
\end{itemize}

In order to develop a timely and appropriate response, different methodologies from the field of control theory are available (see e.g. \cite{Nowzari2016:ieee:CONTROL},
\cite{Bin2020:MULTISHOT:CONTROL}, \cite{casella2020can:Control}, \cite{Kohler2020:mpc:with:FRANK}). We describe them in the following subsections.

\subsection{Optimal Control Theory}

Optimal control theory \cite{Liberzon2012:optimal:control} can be applied to reduce in an effective way the burden of an epidemic \cite{Lenhart2007:OPTIMAL:CONTROL}, \cite[Chapter 9]{Martcheva2015}. The dynamic optimization techniques of the calculus of variations and of optimal control theory provide methods for solving planning problems in continuous time. The solution is a continuous function (or a set of functions) indicating the optimal path to be followed by the variables through time or space \cite{KamienMortonIandSchwartz2012:optimal:control}. 
We present here a common formulation of a continuous dynamical optimization problem \cite[Section 2]{Hartl1995:OPTIMAL:CONTROL}:
\begin{eqnarray}
\min\limits_{x(\cdot),u(\cdot)} &&S(x(T),T)+ \int\limits_0^T F(x(T),u(t),t)dt \nonumber\\
s.t. && x(0)=x_0, \nonumber \\  
&& \dot{x}  =  f(x(t),u(t),t), \nonumber\\
&& g(x(t),u(t),t) \geq 0, \label{ineq:mixed}\\
&& h(x(t),t) \geq 0, \label{ineq:pure}\\
&& a(x(T),T) \geq 0, \label{ineq:Terminal}\\
&& b(x(T),T)  =  0. \label{equ:Terminal}
\end{eqnarray}
In an epidemic control problem $x(t)$ represents the state of the pandemic at time $t$ (for example, in terms of the populations of the different compartments), $u(t)$ is the control action which can be stated in a direct way (intensity of the interventions, number of vaccines, treatments), or in an indirect way (infection rate, immunologic protection, recovery rate). The differential equation $\dot{x}(\cdot)=f(\cdot,\cdot,\cdot)$ represents the epidemic model, constraint (\ref{ineq:mixed}) allows us to incorporate (mixed) constraints on $x(\cdot)$ and $u(\cdot)$ whereas the (pure) constraint (\ref{ineq:pure}) can be used to impose limits on the size of the components of $x(\cdot)$. Finally, (\ref{ineq:Terminal}) and (\ref{equ:Terminal}) are terminal constraints. The question of existence of optimal pairs $(x^*(\cdot),u^*(\cdot))$ was studied in in \cite{Cesari1965:Existence:OPTIMAL:CONTROL} and \cite{Filippov1962:EXISTENCE:OPTIMAL:CONTROL}. See also \cite[Section 3]{Hartl1995:OPTIMAL:CONTROL} and references therein.  

Pontryagin's maximum principle provides necessary conditions that characterize the optimal solutions in the presence of inequality constraints \cite{Liberzon2012:optimal:control}, \cite{Kirk2004:OPTIMAL:CONTROL}. These necessary conditions become sufficient under certain convexity conditions on the objective and constraint functions \cite{Mangasarian1966:OPTIMAL:CONTROL},\cite{Kamien1971:OPTIMAL:CONTROL}. In general, the solution of the optimal problem in the presence of nonlinear dynamics and constraints requires iterative numerical methods to solve the so-called Hamiltonian system, which is a two-point boundary value problem, plus a maximum (minimum) condition of the Hamiltonian (see e.g. \cite[Chapter 6]{Kirk2004:OPTIMAL:CONTROL}).  

We now describe some examples of the use of the optimal control theory in epidemic control. In \cite{Zaman2008:OPTIMAL:CONTROL},
the dynamic optimal vaccination strategy for a SIR epidemic model is described. The optimal solution is obtained using a forward-backward iterative method with a Runge-Kutta fourth-order solver. An example of how to deploy scarce resources for disease control when epidemics occur in different but interconnected regions is presented in \cite{Rowthorn2009:optimal:control}. The authors solve the optimal control problem of minimizing the total level of infection when the control actions are bounded. 

In \cite{Youssef2013} the authors apply Pontryagin's Theorem to obtain an optimal Bang-Bang strategy to minimize the total number of infection cases during the spread of susceptible-infected-recovered SIR epidemics in contact networks. Optimal control theory is employed to design the best policies to control the spread of seasonal and novel A-H1N1 strains in \cite{Prosper2011:OPTIMAL:CONTROL}. An example of the use of optimal control theory to control the present Covid-19 pandemic is presented in \cite{Mandal2020}, where the authors design an optimal strategy, for a five compartmental model, in order to minimize the number of infected cases while minimizing the cost of the non-pharmaceutical interventions.

\subsection{Model Predictive Control}

Model predictive control (MPC) is a methodology that provides optimal solutions to a control decision problem subject to constraints \cite{Camacho2013:MPC}, \cite{rawlings2017:MPC}.  MPC is a receding horizon methodology that involves repeatedly solving a constrained optimization problem, using predictions of future costs, disturbances, and constraints over a moving time horizon. In epidemic control, the aforementioned optimization problem is solved daily, or weekly, in order to decide the optimal control action (for example, the intensity of mitigation interventions, or the optimal allocation of resources). The output of the model  predictive controller is adaptive in the sense that it takes into consideration the latest available information on the state of the pandemic. See, for example \cite{Alleman2020:MPC},  \cite{Kohler2020:mpc:with:FRANK} and \cite{Morato2020:MPC:BRAZIL} for MPC formulations that address the control of the Covid-19 pandemic.

Because of the spatial clustered distribution of an epidemic, it is possible to apply specific control techniques from the field of distributed model predictive control \cite{Maestre2014}, \cite{Christofides2013}. For example, non linear model predictive control can be used to control the epidemics by solely acting upon the individuals’ contact pattern or network \cite{Selley2015}.
Another example of distributed MPC in the control of epidemics is given in  \cite{Kohler2018:MPC:Resource:Allocation},  where the problem of dynamically allocating limited resources (vaccines and antidotes) to control an epidemic spreading process over a network is addressed.  

\subsection{Multi-objective control}

Pareto optimality is used in multi-objective control problems with counter-balanced objectives. For instance, in a counter-balanced bi-objective problem, improving one objective implies to worsen the other one. Pareto optimality is based on the Pareto dominance, which defines that one solution dominates another one iff it is strictly superior in all the objectives. Thus, the goal of the optimization algorithm is to find the Pareto front, which includes all dominant solutions of the control problem. Therefore, there is a set of optimal solutions instead of one optimal solution. The Pareto front is a useful tool for decision-makers that allows to visualize all the possible optimal solutions (for two objectives is a curve, for three objectives a plane, and so forth) and to evaluate the trade-off between different strategies. In the context of epidemic control \cite{Sharomi2017}, Pareto optimality has been used in \cite{Yousefpour2020} in a bi-objective control problem, the goals are related to epidemic measures like the number of cases and economic costs. Thus, the results show that economic and epidemiology goals are counter-balanced. In general, this result can be expected since the security of a population with respect to the spread of a virus can only be guaranteed by increasing the economic costs through contention and mitigation strategies that reduce mobility.     

\subsection{Reinforcement Learning}
In this section, a brief introduction to Reinforcement Learning and its use in the context of epidemic control is presented.

Reinforcement learning is an artificial intelligence technique where an agent learns by actions. For an executed action $a \in A$ in a given state $s \in S$, where $A$ is the set of possible actions in a state $s$ and $S$ is the set of possible states of the agent, the agent will move stochastically to a new state $s^+$. The new state $s^+$ is given by the dynamic model of the environment where the agent performs. Thus, reinforcement learning problems are usually formulated as finite Markov Decision Processes (MDP) \cite{puterman2014markov}. In an MDP, the Markov property indicates that the state includes information about all aspects of the past agent–environment interaction that make a difference for the future. On the transition from one state to another after executing a given action, the agent receives a reward signal $r \in R$. Normally, if an action is positive for the agent's global objective, the reward will be positive. In contrast, bad actions will lead to negative rewards. Thus, the learning procedure is a  trial-and-error search \cite{sutton2018reinforcement}. The agent can know its state $s$ by sensing the environment and the final goal of the agent is to maximize the long term reward received by finding the optimal policy $\pi^*$ in a sequential decision-making scenario. It is important to highlight the long term feature of the agent's goal; otherwise, it will select greedily the best action at each state.  

In \cite{yanez2019towards}, the authors study how to design environments to represent the problem of control epidemics and finding optimal interventions. The main aspects to take into account are:
\begin{itemize}
    \item In the control epidemic problem, the agent represents decision markers such as governments and institutions, and the goal is to find the optimal intervention strategy ($\pi^*$).
    \item The state of the agent is the state of the pandemic measured not only in terms of the size of the different populations of the compartmental model (SIR, SEIR, etc.), but also through the parameters of the underlying epidemic model like infection rate, basic reproduction number, etc. Therefore, the set $S$ represents all possible states of the epidemic model. 
    \item The agent's actions are contention and mitigation techniques established by decision-makers, such as wearing masks, social distance, contact tracing, closing schools, lockdowns, etc.
    \item The reward signal is related to the decision maker' goals. For instance, if the goal is to reduce the total number of deaths, the reward can be related with the rate of deaths at each time step by a simple function growing with the decrease of the rate of deaths.
    \item The policy is a function that indicates an action $a \in A$ to perform for each state $s \in S$, $\pi:S\rightarrow  A$. 
\end{itemize}

From the optimization point of view, the problem to solve is to find the optimal policy $\pi$ that selects at each state $s$ the optimal action $a$  to obtain the maximum expected reward after
a number of decision $t \in T$ (long-term benefit), $\pi^{*} = argmax_{\pi}E[\sum_{t=1}^{T}r_{t}]$. There are different methods to solve the optimization problem, such as dynamic programming and heuristic approaches with exploration and exploitation capabilities. Dynamic optimization requires a dynamic model of the environment (pandemic) in order to access to a complete MDP representation of the disease. As examples of this case, see \cite{yaesoubi2016identifying}, where an approximate dynamic policy aimed at optimizing the population's net health benefit for H1N1 influenza is proposed, and \cite{yaesoubi2013identifying}, where it is applied to tuberculosis epidemics. If a dynamical model is not available or it is too complex, there are free-model solutions like temporal difference learning \cite{sutton2018reinforcement} and Q-learning \cite{watkins1992q}. In \cite{libin2018bayesian}, the authors model the problem of selecting optimal strategies for influenza as k-armed bandit problem, which is well-known in the reinforcement learning community. Bayesian best-arm identification algorithms \cite{libin2019bayesian}, such as top-two Thompson sampling and BayesGap, are evaluated to solve the problem.

There are some limitations of reinforcement learning approaches:
\begin{itemize}
    \item It requires a training process to obtain the optimal policy. The duration of the training process depends on the complexity of the problem. Therefore, a realistic epidemic simulation environment is crucial to obtain the optimal policy. In addition, since the process requires trial-error steps and heuristics, it cannot be trained on real-time.
    \item The number of states and the observability of the environment can be limiting factors.
    \item The definition of a suitable reward signal is also sometimes challenging. 
\end{itemize}

Reinforcement learning approaches have also been combined with deep learning techniques in Deep Reinforcement Learning methods \cite{mnih2015human}. In these approaches, the power of deep learning algorithms for data representation is used to obtain the optimal policy. For instance, in \cite{libin2020deep}, a deep learning approach is used to learn
prevention strategies in the context of influenza epidemics. 


\section{Conclusions}\label{sec:Conclusions}

This document presents a roadmap for controlling Covid-19 pandemic from a 3M data-driven perspective: Monitoring, Modelling, and Making decisions. A holistic approach is required to efficiently content and mitigate the impact of the pandemic. We have highlighted the interplay between data science, epidemiology, and control theory in a data-driven approach to address the different challenges raised by the pandemic. 

Methodologies and approaches proposed for previous epidemics and current Covid-19 have been reviewed. Although the literature is large and many approaches have been studied in the past, further research is still necessary. The implementation of effective control strategies to mitigate the pandemic is difficult because of different reasons:  i) the significant non-symptomatic transmission of Covid-19, ii) the uncertainty on some crucial parameters characterizing the spread, iii) the difficulties in assessing the quantitative effect of the mitigation interventions, iv) the impossibility of obtaining a prompt evaluation of the effect of the implemented intervention. 

The first step for the modelling of different aspects of the pandemic is the processing of the available raw data to obtain consolidated time-series. Different methodologies from the field of Data-Reconciliation, Data-Fusion, Clustering Methods and Signal Processing have been reviewed. In order to obtain prediction models, which are crucial for the decision-making process, we provide an enumeration of techniques from epidemiology and machine learning. We describe the most relevant modelling and forecasting approaches focusing on the adjustment of the prediction models to the available data, model selection and validation processes. 

The most plausible scenario may involve recurring epidemic waves interspersed with periods of low-level transmission. Thus, different surveillance systems able to detect, or anticipate, possible recurring epidemic waves are surveyed. These systems enable an immediate response that reduces the potential burden of the outbreak. We also analyze different techniques for the implementation of surveillance and monitoring systems, enabling a prompt response in case of a secondary epidemic wave. Different methods from control theory can be applied to provide an optimal, robust and adaptive response to the time-varying incidence of the epidemic. These methods can be applied to the optimal allocation of resources, and to the determination of trigger control schemes that increase or decrease the intensity of the interventions.  
We review the control-theory literature in the context of analysis and design of feedback structures leading to efficient control of an epidemic.
Besides, we also mention some techniques from distributed model predictive control that can be applied to address the temporal and spatial dimension of the spread of Covid-19.

\subsection{Updates and Contributors}

To keep pace with the pandemic, this paper will be updated regularly in arXiv\footnote{https://arxiv.org/} during the Covid-19 pandemic. We thank feedback and collaboration from the community to enrich and update the paper. Contributors can contact the CONCO-Team via e-mail: conco.team@gmail.com.    


\bibliography{Bib_Survey_Data_Driven}

\end{document}